\documentclass[%
 reprint,
 amsmath,amssymb,
 aps,
 prb,
]{revtex4-2}
\usepackage{graphicx}
\usepackage{dcolumn}
\usepackage{bm}

\usepackage{booktabs,makecell,multirow,braket,bm,amssymb,pifont}
\usepackage{threeparttable}
\usepackage{hyperref}
\hypersetup{colorlinks,
	linkcolor=blue,%
  filecolor=blue,      
  urlcolor=blue,
  anchorcolor=blue,
	citecolor=blue}
\usepackage[mathlines]{lineno}

\begin{document}

\preprint{APS/123-QED}

\title{Localized State-Induced Enhanced Emission in Perovskite Nanocrystals}

\author{Feilong Wang}
\author{Qiongrong Ou}%
\author{Shuyu Zhang}
\email{zhangshuyu@fudan.edu.cn}

\affiliation{%
	Institute for Electric Light Sources, School of Information Science and Technology, Fudan University, Shanghai 200433, China.}%

\date{\today}

\begin{abstract}
	The luminescent mechanism of nano-sized materials with indirect bandgap and parity-forbidden transition has always been a critical issue in breaking through bottlenecks in traditional luminescent materials. The lack of understanding has resulted in great disputes regarding the origin of fluorescence in weak-transition nanocrystals. This paper proposes a new physical luminescence model, named as localized state-induced enhanced emission, i.e., localized processes induced by size-limitation or partial doping, to explain the anomalous luminescence from non-luminescent state to luminescent state in various perovskite nanocrystals with indirect/direct bandgap or parity-forbidden transition. These findings provide a theoretical viewpoint to design efficient lead-free perovskite NCs and promote the development of fast optical transitions in luminescent materials.
\end{abstract}

\maketitle


\section{\label{sec:level1} INTRODUCTION}

Nanocrystals (NCs) have emerged as a promising luminescent material, playing a significant role in the fields of luminous displays\cite{yangRecentAdvancesQuantum2019,quElectroluminescenceNanocrystalsUm2022}, laser devices\cite{rohOpticallyPumpedColloidalquantumdot2020,jungProspectsChallengesColloidal2021}, and biochemical sensing\cite{freemanOpticalMolecularSensing2012,fery-forguesFluorescentOrganicNanocrystals2013a}.
Consequently, they have greatly advanced the development of luminescence physics. Although basic phenomenological physical luminescence models have been established over the past few decades, including emission from quantum confinement\cite{zhuPhotoluminescenceMechanismGraphene2017,dongPreciseControlQuantum2018a}, luminescence due to defect or doping\cite{zhangDefectrelatedLuminescentMaterials2012,zhouHighEfficiencyFastRadiativeBlueEmitting2022}, self-trapped exciton luminescence\cite{beninHighlyEmissiveSelfTrapped2018,jingPhotoluminescenceSingletTriplet2021}, and aggregation-induced emission\cite{luoAggregationinducedEmission1methyl12001,zhangAggregateScienceStructures2020}, disputes still exist regarding the phase transition in indirect bandgap semiconductors from a non-luminescent to a luminescent state and in direct bandgap semiconductors from a forbidden transition to an allowable one.

Effective luminescence with high photoluminescence quantum yield (PLQY) has been observed in nanocrystalline systems corresponding bulk materials that typically exhibit difficulty in emitting light, such as silicon NCs\cite{pringleBrightSiliconNanocrystals2020a} and lead-free double perovskite Cs\textsubscript{2}AgBiX\textsubscript{6} (X=Cl, Br) NCs\cite{duBandgapEngineeringLeadFree2017}. The origin of luminescence in silicon NCs has been elucidated recently with a new physical luminescence model called Localized State-Induced Enhanced Emission (LIEE)\cite{wangEnhancementIntrinsicOptical2022}, which successfully explained the luminescence in silicon NCs and the fast optical transition resulting from size reduction. However, research has yet to demonstrate the universality of this model, and the fluorescence mechanism in indirect-bandgap Cs\textsubscript{2}AgBiX\textsubscript{6} NCs remains elusive\cite{duBandgapEngineeringLeadFree2017,deyTransferDirectIndirect2020}. 

Despite the potential of NCs as luminescent materials, there remains a lack of precise understanding of the anomalous luminescence mechanisms and how to improve their luminous performance. Luo et al. suggested that the change in electron wavefunction and increased electron-hole (e-h) pair overlap, resulting from Jahn-Teller distortion of the AgCl\textsubscript{6} octahedron in sodium-doped Cs\textsubscript{2}AgInCl\textsubscript{6}, led to an enlarged transition dipole moment (TDM)\cite{luoEfficientStableEmission2018}. However, the impact of the surface effects on e-h pairs was not taken into account in (sodium-doped) Cs\textsubscript{2}AgInCl\textsubscript{6} NCs. In a related study, the yellow emission in silver-doped Cs\textsubscript{2}NaInCl\textsubscript{6} was ascribed to bright self-trapped excitons\cite{hanLeadFreeSodiumIndium2019} without consideration of the surface scattering (electron scattering at nanocrystalline surface).
To date, few reports have been able to pinpoint the origins of luminescence in double perovskite NCs or clarify the enhanced emission caused by partial doping. In short, one of the primary obstacles in the development of lead-free double perovskites is the absence of a comprehensive understanding of the luminescence phenomena in systems with intrinsically low luminescent properties.

In this manuscript, we presents a new approach to explore the luminescence properties of perovskite NCs with poor optical transitions using LIEE due to the scattering of electron at nanocrystalline surface, a method previously applied to silicon NCs. We began by calculating the wavefunction distribution of direct and indirect bandgap perovskites in real space. Then reconstructed the e-h pairs in real space under the single particle excitation approximation and demonstrated that LIEE can produce anomalous luminescence in perovskite NCs without the need for electron-phonon coupling or Jahn–Teller distortion effects. Our results provide a theoretical foundation for designing efficient lead-free double perovskites NCs and may advance the development of fast optical transitions in luminescent materials.

\section{COMPUTATIONAL DETAILS}

All-electron full-potential linearised augmented-plane wave (LAPW)
method was employed with ELK code\cite{ElkCode} in bulk-materials calculation. All
structures were optimized with Perdew-Burke-Ernzerhof (PBE)\cite{perdewGeneralizedGradientApproximation1996} functional
and a 6$\times$6$\times$6 $k$-points grid was adopted.
$r\textsuperscript{2}$SCAN\cite{furnessAccurateNumericallyEfficient2020} was employed to obtain energy band and Kohn-Sham (KS) wavefunction.

All ground-state calculations of clusters were performed in CP2K.\cite{kuhneCP2KElectronicStructure2020}
Gaussian plane waves method (GPW) in the QUICKSTEP\cite{vandevondeleQuickstepFastAccurate2005} module was employed
with a double-zeta DZVP-MOLOPT-SR-GTH basis set\cite{vandevondeleGaussianBasisSets2007a} and norm-conserving
GTH-PBE\cite{goedeckerSeparableDualspaceGaussian1996,krackPseudopotentialsKrOptimized2005}
pseudopotential. A 500-Ry cutoff energy was applied for
planewave expansion, and each Gaussian was mapped onto a 55-Ry cutoff
grid. Dispersion with D3\cite{grimmeConsistentAccurateInitio2010b} was considered in all calculations. All
structures were placed in larger non-periodic boxes (thus in cluster model) and then optimized for minimal energy with
Broyden-Fletcher-Goldfarb-Shanno (BFGS) algorithm in PBE\cite{perdewGeneralizedGradientApproximation1996c} level.

All excited state calculation were performed in TDA-TDDFT (
Tamm-Dancoff approximated time-dependent density functional theory).
Configuration coefficients greater than 0.001 were printed out for e-h pairs analysis.
Multiwfn\cite{luMultiwfnMultifunctionalWavefunction2012} was employed to conduct e-h pairs analysis.

In TDDFT framework, all KS orbital are involved in electron excited
state. The excited state wavefunction is

\begin{equation}
	\Psi_{TD}=\sum_{i \rightarrow a} \omega _{i}^{a}
	\Phi _{i}^{a} + \sum_{a \rightarrow i} \omega _{i}^{'a} \Phi _{i}^{a}
\end{equation}

$\Phi _{i}^{a}$ is the configuration state
wavefunction corresponding to moving an electron from originally
occupied $i$-th MO to virtual $a$-th MO. $\omega _{i}^{a}$ and $\omega _{i}^{'a}$ are configuration coefficient of
excitation and de-excitation, respectively, limited by normalizing
conditions. In TDA-TDDFT, the de-excitation term disappears and the hole and electron density can be perfectly
defined as\cite{liuSphybridizedAllcarboatomicRing2020}

\begin{equation}
	\rho_{h}(\mathbf{r})=\sum_{i \rightarrow a} {(\omega _{i}^{a})}^2
	\varphi_{i}(\mathbf{r}) \varphi_{i}(\mathbf{r}) + \sum_{i \rightarrow a}
	\sum_{j\ne i \rightarrow a} \omega _{i}^{a} \omega _{j}^{a} \varphi_{i}(\mathbf{r})
	\varphi_{j}(\mathbf{r})
\end{equation}

\begin{equation}
	\rho_{e}(\mathbf{r})=\sum_{i \rightarrow a} {(\omega _{i}^{a})}^2
	\varphi_{a}(\mathbf{r}) \varphi_{a}(\mathbf{r}) + \sum_{i \rightarrow a}
	\sum_{ i \rightarrow b\ne a} \omega _{i}^{a} \omega _{i}^{b} \varphi_{a}(\mathbf{r})
	\varphi_{b}(\mathbf{r})
\end{equation}

$\varphi_{i}$ and $\varphi_{j}$ are occupied
molecular orbitals, while $\varphi_{a}$ and $\varphi_{b}$
are virtual orbitals.

The transition density in real space can be writen as
\begin{equation}
	\mathbf{T}_{\rho}( \mathbf{r})=\sum_{i \rightarrow a}\omega _{i}^{a}
	\psi_i( \mathbf{r}) \psi_a( \mathbf{r})
\end{equation}

The integral in the whole space after multiplying it with the coordinate $\mathbf{r}$
\begin{equation}
	\mathbf{d}= -\mathbf{r} \cdot \mathbf{T}_{\rho}( \mathbf{r})
\end{equation}
is the transition dipole moment (TDM) $\mathbf{d}$.

\begin{figure*}
	\centering
	\includegraphics[width=\linewidth,scale=0.5]{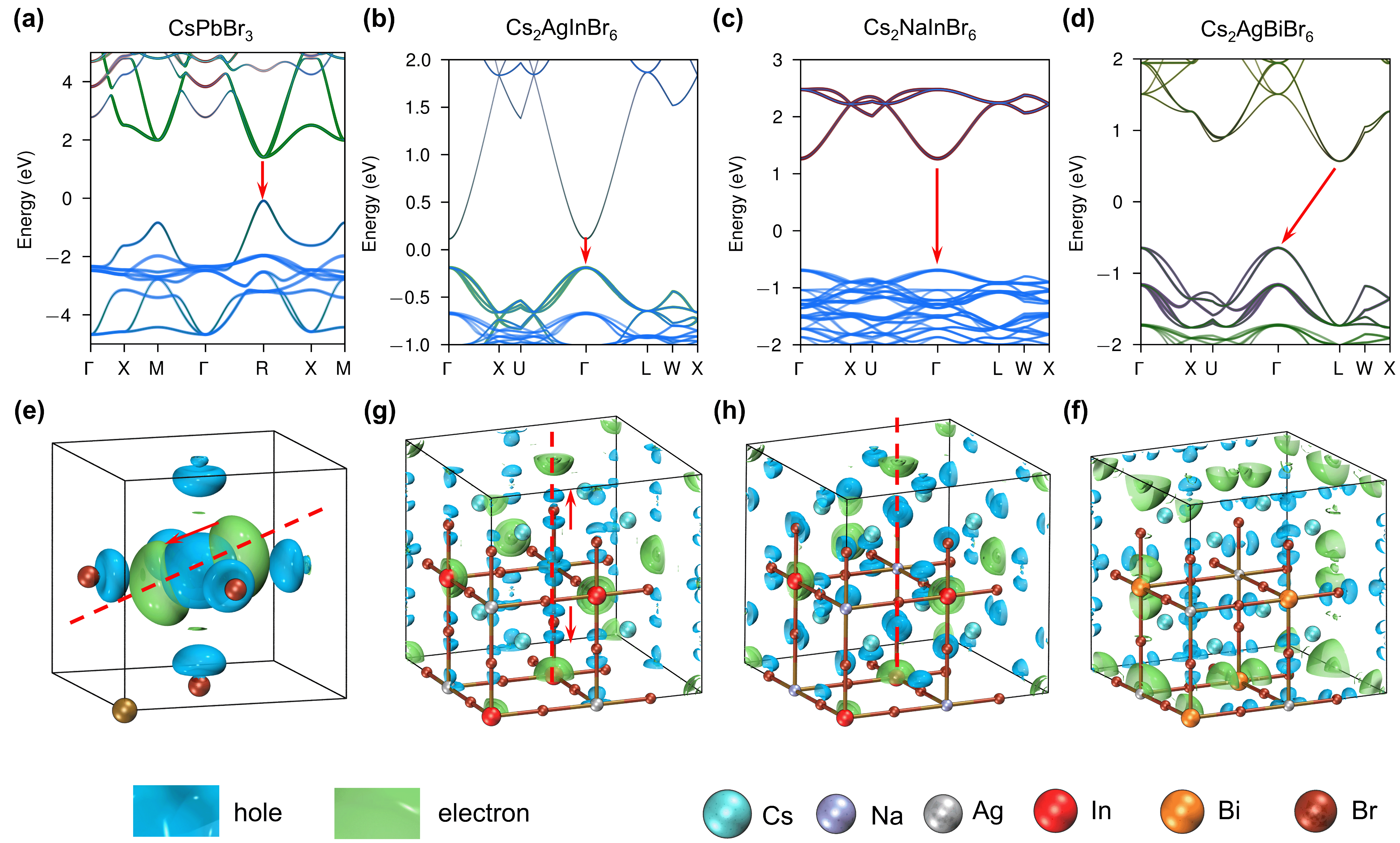}
	\caption{\label{fig:f1} Energy band diagram and isosurface of
		electron-hole in (a) and (e) CsPbBr\textsubscript{3},
		(b) and (f) Cs\textsubscript{2}AgInBr\textsubscript{6},
		(c) and (g) Cs\textsubscript{2}NaInBr\textsubscript{6} and
		(d) and (h) Cs\textsubscript{2}AgInBr\textsubscript{6}. Red arrows in (g) show two
		antipodal transition dipole sub-vector due to crystal symmetry, .ie. ,
		the electron and hole possess the same parity. The red dotted lines facilitate observation.
	}
\end{figure*}

\begin{table*}
	\centering
	\caption{A brief summary of transition condition in bulk perovskite. Orbital symmetry dictates the atomic optical transition (the partial TDM ``$\uparrow$'' or ``$\downarrow$''). The symmetry of these partial TDMs, in turn, defines the overall TDM for the system. The orientation of the orbitals then governs the degree of overlap.}
	\label{tab:table_1}
	\begin{threeparttable}
	\begin{tabular}{ccccccc}
		\toprule
		\multicolumn{1}{c}{\multirow{2}*{kind}}                        & \multicolumn{1}{c}{\multirow{2}*{overlap(a. u.)}}
		                                                               & \multicolumn{2}{c}{symmetry}                      & \multicolumn{1}{c}{\multirow{2}*{orientation}}
		                                                               & \multicolumn{1}{c}{\multirow{2}*{$f_\mathrm{osc}$ (a.u.)}}      & \multicolumn{1}{c}{\multirow{2}*{emit}}                                             \\
		\multicolumn{1}{c}{}                                           &                                                   & orbital                                        & partial TDM                      & \\
		\hline
		\multicolumn{1}{c}{CsPbBr\textsubscript{3}}                    & 7.295                                             & \checkmark \checkmark \checkmark               & \checkmark \checkmark \checkmark
		                                                               & \checkmark \checkmark \checkmark                  & $2.7$                                          & excellent                          \\
		\multicolumn{1}{c}{Cs\textsubscript{2}AgInBr\textsubscript{6}} & 3.307                                             & \checkmark \checkmark \checkmark               &
		\ding{55}                                                      & \checkmark \checkmark                             & $3.6 \times 10^{-22}$                          & no                                 \\
		\multicolumn{1}{c}{Cs\textsubscript{2}NaInBr\textsubscript{6}} & 2.976                                             & \checkmark                                     & \checkmark
		                                                               & \ding{55}                                         & $9.1 \times 10^{-17}$                          & no                                 \\
		\multicolumn{1}{c}{Cs\textsubscript{2}AgBiBr\textsubscript{6}} & 3.119                                             & \checkmark                                     & \checkmark
		                                                               & \ding{55}                                         & $1.3 \times 10^{-5}$ (regardless of momentum)  & no                                 \\
		\bottomrule
	\end{tabular}
	\begin{tablenotes}
		\item[]\checkmark \ favorable degree of luminescence, \ding{55} not conducive to luminescence.
	\end{tablenotes}
\end{threeparttable}
\end{table*}

\begin{figure}
	\centering
	\includegraphics[width=\linewidth]{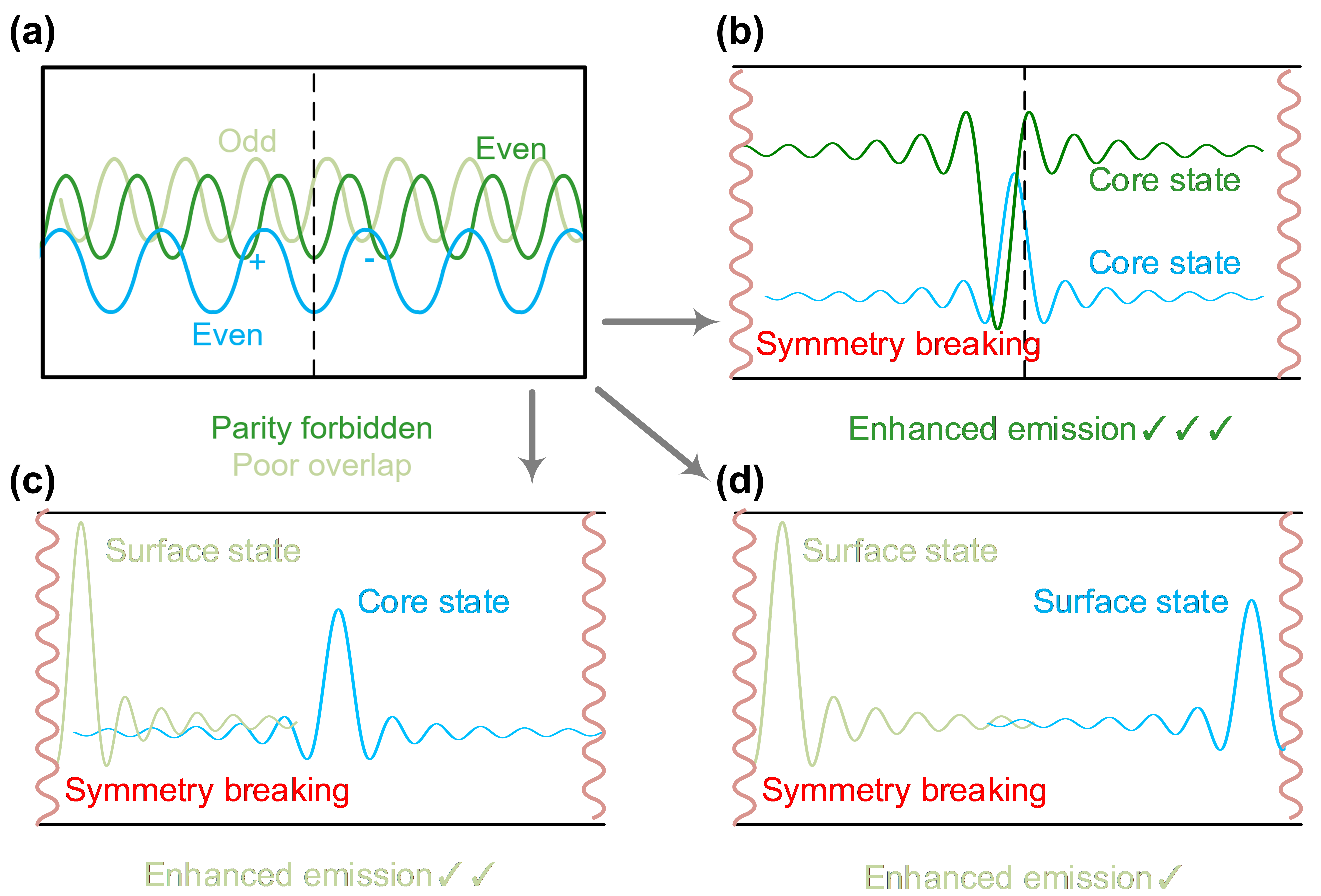}
	\caption{\label{fig:f2} A brief illustration in the formation of localized states.
		(a) In periodic bulk, e-h wavefunctions are described as periodic Bloch functions.
		The ``even-to-even" transition due to the cancellation of the two partial TDM ``$\downarrow \uparrow$"
		and the weak ``even-to-odd" transition with little overlap are both shown.
		(b) In a good nano-structured ``cavity", the e-h are highly concentrated in some areas.
		This situation is extremely beneficial to the enhancement of emission. However,
		(c) surface-to-core localized states and  (d) surface-to-surface localized states
		could be arised limited by the surface and overlap of the e-h is decreased sharply.
		Therefore, their enhancement in luminescence is much limited.
		Nonetheless, emission in nanomaterials is more efficient than bulk for media difficult to emit.
		It should be noted that the TDM strongly dependents on the size of nanomaterials in the latter two cases.
	}
\end{figure}

\begin{figure*}
	\centering
	\includegraphics[width=\linewidth,scale=0.5]{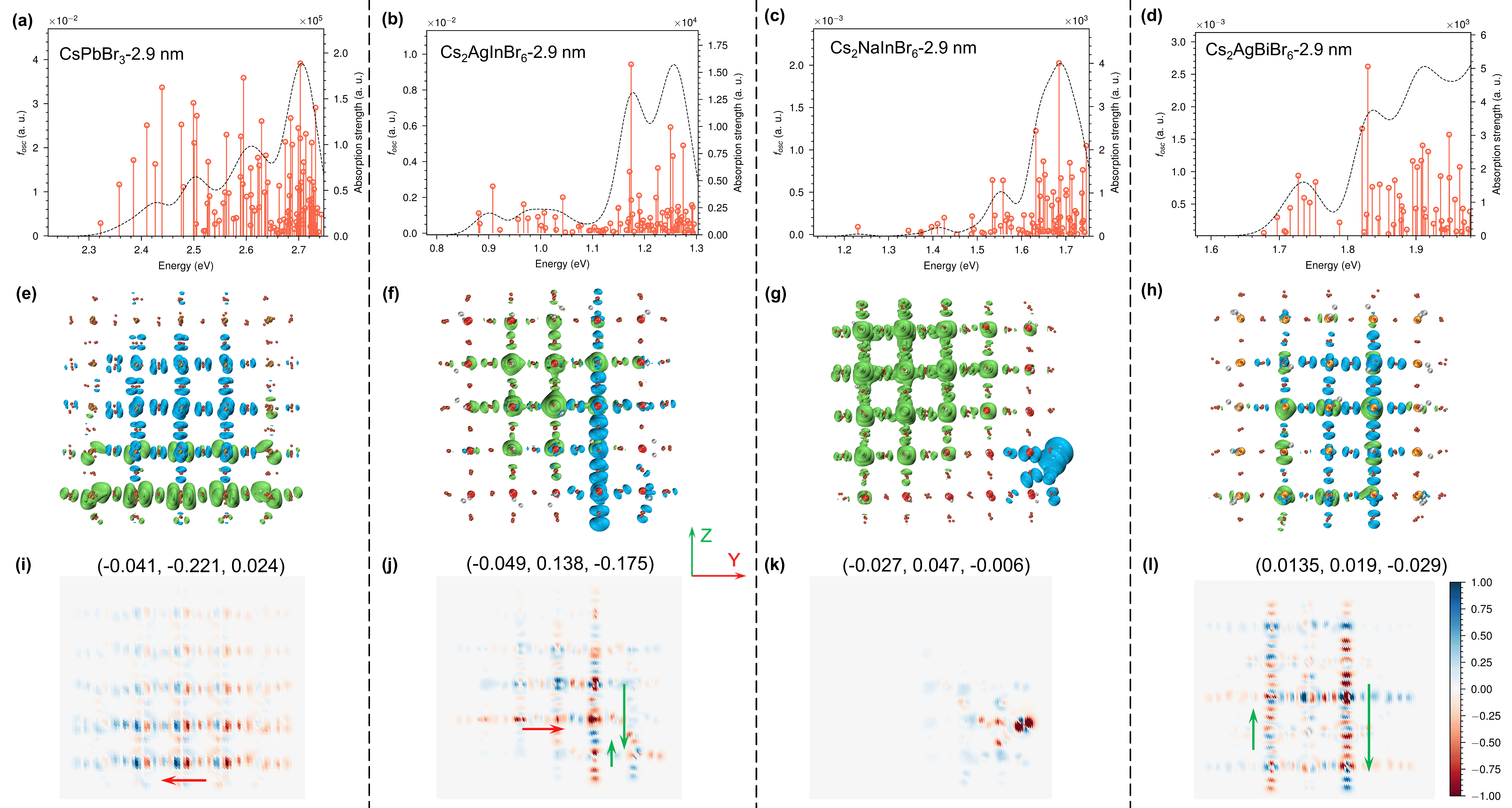}
	\caption{\label{fig:f3} Optical transition in perovskite nanocrystals with (a)-(d) absorbing oscillator strength.
	(e)-(h) e-h distribution of the lowest transition state (blue isosurface for hole, green isosurface for electron).
	And (i)-(l) the YZ-plane projection of transition density with TDM above the figure. The arrows represent transition dipole sub-vectors which the opposite direction would partially cancel out.}
\end{figure*}

\section{RESULTS AND DISCUSSION}
\subsection{E-h pair distribution in bulk perovskite}
Firstly, we have calculated energy band and obtained Kohn-Sham wavefunction for both direct and indirect bulk-perovskite structures (Fig. \ref{fig:f1}). In CsPbBr\textsubscript{3}, the hole is made up of the \emph{p}-orbital of bromine and the 6\emph{s}-orbital of lead, while the electron is mainly located on the 6\emph{p}-orbital of lead. This transition satisfies the quantum transition selection rule and possesses an excellent luminous rate, as electrons and holes in the same real space greatly increase the transition probability according to the Dirac-Fermi's golden rule. The TDM is maximal due to opposing orbital symmetry, as shown in TAB. \ref{tab:table_1}. Unfortunately, the lead-free perovskite structures Cs\textsubscript{2}AgInBr\textsubscript{6}  Cs\textsubscript{2}NaInBr\textsubscript{6} with direct bandgaps exhibit poor luminescence. The holes are provided by the \emph{p}-orbital of bromine (and the \emph{$d_{z^2}$}-orbital of silver in Cs\textsubscript{2}AgInBr\textsubscript{6}), while indium (with silver) contributes \emph{s}-orbital to the electrons (Fig. \ref{fig:f1}g and h). From the perspective of the quantum transition selection rule, the silver-to-indium (silver) transition is strictly prohibited, but the bromine-to-indium (silver) (odd-to-even parity) transition can occur. In Cs\textsubscript{2}AgInBr\textsubscript{6}, atoms along the In-Br-Ag-Br-In axis (the red line in Fig. \ref{fig:f1}b) contributes to the TDM presenting an extremely symmetrical transition density ``$+--+$" or ``$-++-$". That means if the ``$+$" represents In($s$)$\rightarrow$Br($p$), ``$-$" represents Br($p$)$\rightarrow$Ag($d$) and the ``$+\rightarrow-$" is the partial TDM. The TDM should add the two partial TDM, ``$+\rightarrow-$"($\downarrow$) and ``$-\rightarrow+$"($\uparrow$) in Fig. \ref{fig:f1} (two reds arrows). Thus, opposite partial TDM in Fig. \ref{fig:f1} (two reds arrows) are simultaneously contributing to the TDM. If the system only have one partial TDM mentioned above, The optical transition clearly occurrs. The cancellation of opposite direction of the two partial TDM results in a zero TDM, known as a crystal symmetry-forbidden transition. Another interesting forbidden mechanism is presented in Cs\textsubscript{2}NaInBr\textsubscript{6}. The absence of coupling interaction between sodium and bromine induces a nodal plane of bromine \emph{p}-orbital that is parallel to the indium-bromide bond, meaning that the hole and electron cannot effectively overlap (Cs\textsubscript{2}NaInBr\textsubscript{6} vs. Cs\textsubscript{2}AgInBr\textsubscript{6}, $2.976$ vs. $3.307$ in TAB. \ref{tab:table_1}). Thus, an extremely limited transition probability arises in Cs\textsubscript{2}NaInBr\textsubscript{6}.

In another lead-free perovskite, Cs\textsubscript{2}AgBiBr\textsubscript{6} is considered poorly luminescent due to its indirect bandgap. The silver 4\emph{$d_{z^2}$}-orbital, bismuth 6\emph{p}-orbital, bromine 4\emph{p}-orbital, and silver 5\emph{s}-orbital can satisfy the necessary symmetry requirements. However, the perpendicular orientation of orbitals and mismatching momentum results in repulsion between holes and electrons, hindering luminescence.

Indeed, to determine luminescence probability, it is necessary to consider the electric dipole transition, $\bra{j} \mathbf{r} \ket{i}$, which involves the vector sum of partial TDMs, crystal orbital, or molecular orbital symmetry and effective electron-hole overlap. Additionally, the transition must satisfy the selection rule ($\Delta{l}=\pm 1$).

\subsection{Origin of the anomalous emission in NCs}
Obviously, the primary approach to obtain emission in poor-luminescence
materials is to change the e-h spatial distribution, so as to overcome the luminescent restrictions
as list in TAB. \ref{tab:table_1}. Localized states provide an effective means to change the non-luminescent situation.
In the past few decades, despite unclear description to abnormal emission in space-limited nano-system,
researchers have observed luminescence in silicon through porous silicon \cite{wolkinElectronicStatesLuminescence1999}, quantum well \cite{canhamSiliconQuantumWire1990,luQuantumConfinementLight1995}, and quantum dot \cite{pavesiOpticalGainSilicon2000}.
A recent study suggests that localized states can modify the spatial orientation of electron-hole pairs and produce anomalous emission. In nanomaterials, the electron-hole wavefunction does not satisfy the perfect Bloch function and oscillates in a "cavity" (Fig. \ref{fig:f2}).
Influenced by relaxation of surface atoms, which is related to the surface dielectric layer such as ligands and solutions, the wavefunction evolves into two forms of localized states - core states and surface states - that are no longer symmetrical and highly localized. Therefore, global symmetry is not the principal consideration for efficient emission. Instead, the orientation of the electron-hole pairs overlap and the symmetry of partial TDM are more crucial.

Similar to silicon, can space-limited double perovskite emit? In order to
further elucidate this problem, we calculated the optical transitions of all nanocrystals (Fig. \ref{fig:f3}). All double perovskite NCs present the first exciton absorption peak
with medium oscillator strength (10\textsuperscript{-2}-10\textsuperscript{-4}) that is significantly greater than that observed in bulk materials.This observation suggests that double perovskite NCs possess the capacity to emit light similarly to silicon NCs.
To gain a more intuitive understanding of the localized states, we first generated an isosurface graph of electron-hole pairs, and then projected transition density onto the YZ-plane, as shown in Fig. \ref{fig:f3}. Nearly all transitions were localized and capable of changing the spatial distribution of electron-hole pairs, thereby causing the transition from a non-luminescent (or weakly luminescent) state to a strong optical transition in all nanocrystals.

In some nanocrystals, the excited state contributed by surface atoms is lower than that contributed by atoms inside nanocrystals, which causes electrons to form surface-localized states that are unsuitable for emission. In CsPbBr\textsubscript{3} NCs, the oscillator strength of the first excitated peak is only 0.03 (Fig. \ref{fig:f3}a), which is significantly smaller than the value of 2.7 observed in bulk materials (TAB. \ref{tab:table_1}). Observing the e-h wavefunctions, it is evident that the overlap of e-h pairs is decrescent (Fig. \ref{fig:f3}e), and surface-localized electrons and core-localized holes could cause this reduction. A more conservative case can be seen in Cs\textsubscript{2}NaInBr\textsubscript{6} NCs (Fig. \ref{fig:f3}g), where the e-h pairs tend to be completely separate along the two sides of the surface, resulting in a much smaller oscillator strength of ($10^{-4}$). Nevertheless, it has been greatly improved by approximately $10^{12}$ (Fig. \ref{fig:f3}c) compared to Cs\textsubscript{2}NaInBr\textsubscript{6} bulk (TAB. \ref{tab:table_1}).

In Cs\textsubscript{2}AgInBr\textsubscript{6} and Cs\textsubscript{2}AgBiBr\textsubscript{6}, core-to-core states (shown in Fig. \ref{fig:f2}b) are generated, as depicted in Fig. \ref{fig:f3}f and h. In the former material, localized states have broken the transition symmetry, which means the cancellation of partial TDM (vector addition of ``$\uparrow$'' and ``$\downarrow$'') is avoided, resulting in a $10^{19}$-fold (Fig. \ref{fig:f3}b vs. TAB. \ref{tab:table_1}) improvement in oscillator strength. This material inherits the direct bandgap and high overlap characteristics of the bulk material. On the other hand, for the indirect-bandgap Cs\textsubscript{2}AgBiBr\textsubscript{6}, the main factor limiting the oscillator strength is the effective partial TDM. The optical transition between the $p$-orbital of bismuth and the $p$-orbital of bromine is prevented, and there is only one transition channel—the $s$-orbital of silver and the $p$-orbital of bromine (``$\downarrow$''). However, induced by the localized state, the partial TDM vector between the silver and the surrounding four bromine atoms is not completely symmetrical. Therefore, the vector sum of partial TDM is a non-zero value, albeit an extremely limited one.

\subsection{To precisely control the localized state for efficient luminescence}
Localized states arising from quantum confinement exhibit limited robustness, as depicted in Fig. \ref{fig:f3}. Influenced by the atoms relaxation on surface, the elimination of atomic degeneracy, and the presence of surrounding ligands, the localized state tends to move in the opposite direction of the desired. To ensure that localized states in NCs contribute positively to luminescence, more accurate methods beyond intrinsic quantum confinement are necessary. It meets two essential criteria: 1) creating a scattering surface for localization, and 2) positioning the localized energy level in the lowest unoccupied orbital. Proper doping serves as a simple and effective strategy. Using CsPbBr\textsubscript{3} NCs as an illustration, it is observed that localized states have a negative impact, as shown in Fig. \ref{fig:f3}e. 

\begin{figure}
	\centering
	\includegraphics[width=\linewidth]{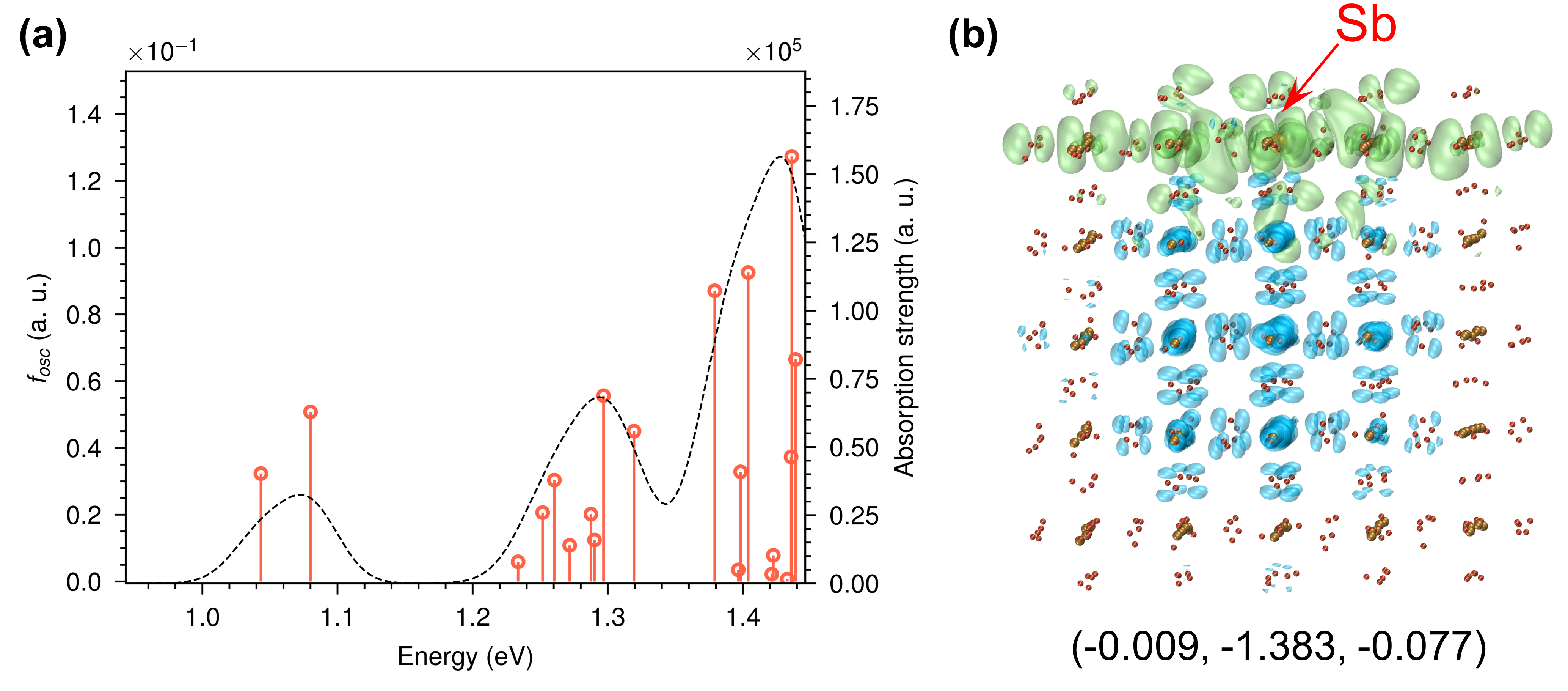}
	\caption{\label{fig:f4} Sb\textsuperscript{3+}-doped CsPbBr\textsubscript{3} with (a) enhanced absorption sepectrum and
	(b)	e-h pairs spatial distribution of the lowest excited states (the TDM vector is marked). 
}
\end{figure}

In Fig. \ref{fig:f4}, Sb\textsuperscript{3+}-doped CsPbBr\textsubscript{3} NCs present three distinct features, consistent with recent experimental results \cite{zhouHighEfficiencyFastRadiativeBlueEmitting2022}: 1) A spectral redshift attributed to the lower Sb\textsuperscript{3+} $5p$ orbital energy level compared to the Pb\textsuperscript{2+} $6p$ orbital, enabling the formation of the lowest localized excited state around the Sb\textsuperscript{3+} atom. 2) Increased radiative recombination rate achieved through a redistribution of electron distribution from Fig. \ref{fig:f2}c to Fig. \ref{fig:f2}b. Notably, Sb\textsuperscript{3+} does not directly involved in luminescence and help to scatter electron to form the Fig. \ref{fig:f2}b model. 3) The redshift narrower emission spectrum results from a trapped state, which widens the energy gap between the several lowest excited states and makes it difficult for thermal transition from the lowest excited states (emitting center) to higher excited states, thereby preventing spectral broadening.

\section{CONCLUSION}
In summary, we have investigated localized state-enhanced emisssion in different types of perovskite NCs. The results indicate that LIEE can control the distribution of the e-h wavefunction and lead to an anomalous luminescence in different kinds of poor luminescent perovskite NCs, via destroying the lattice symmetry, .ie., surface scattering, surface doping, etc. These findings provide a theoretical viewpoint to design efficient lead-free double perovskite NCs and promote the development of fast optical transitions in luminescent materials.

\section{AUTHOR CONTRIBUTIONS}
Conceptualization, Data curation, Investigation, Methodology, Software, Visualization, Writing – original draft: F.W.; Funding acquisition, Project administration, Supervision: S.Z.; Resources: S.Z. and Q.O.; Writing – review \& editing: F.W. and S.Z..

\begin{acknowledgments}
	This work is supported by Science and Technology commission of
	Shanghai Municipality (21ZR1408800), and National Natural Science Foundation of China (11975081).
\end{acknowledgments}

\nocite{*}

\bibliography{apssamp}

\begin{thebibliography}{43}%
\makeatletter
\providecommand \@ifxundefined [1]{%
 \@ifx{#1\undefined}
}%
\providecommand \@ifnum [1]{%
 \ifnum #1\expandafter \@firstoftwo
 \else \expandafter \@secondoftwo
 \fi
}%
\providecommand \@ifx [1]{%
 \ifx #1\expandafter \@firstoftwo
 \else \expandafter \@secondoftwo
 \fi
}%
\providecommand \natexlab [1]{#1}%
\providecommand \enquote  [1]{``#1''}%
\providecommand \bibnamefont  [1]{#1}%
\providecommand \bibfnamefont [1]{#1}%
\providecommand \citenamefont [1]{#1}%
\providecommand \href@noop [0]{\@secondoftwo}%
\providecommand \href [0]{\begingroup \@sanitize@url \@href}%
\providecommand \@href[1]{\@@startlink{#1}\@@href}%
\providecommand \@@href[1]{\endgroup#1\@@endlink}%
\providecommand \@sanitize@url [0]{\catcode `\\12\catcode `\$12\catcode `\&12\catcode `\#12\catcode `\^12\catcode `\_12\catcode `\%12\relax}%
\providecommand \@@startlink[1]{}%
\providecommand \@@endlink[0]{}%
\providecommand \url  [0]{\begingroup\@sanitize@url \@url }%
\providecommand \@url [1]{\endgroup\@href {#1}{\urlprefix }}%
\providecommand \urlprefix  [0]{URL }%
\providecommand \Eprint [0]{\href }%
\providecommand \doibase [0]{https://doi.org/}%
\providecommand \selectlanguage [0]{\@gobble}%
\providecommand \bibinfo  [0]{\@secondoftwo}%
\providecommand \bibfield  [0]{\@secondoftwo}%
\providecommand \translation [1]{[#1]}%
\providecommand \BibitemOpen [0]{}%
\providecommand \bibitemStop [0]{}%
\providecommand \bibitemNoStop [0]{.\EOS\space}%
\providecommand \EOS [0]{\spacefactor3000\relax}%
\providecommand \BibitemShut  [1]{\csname bibitem#1\endcsname}%
\let\auto@bib@innerbib\@empty
\bibitem [{\citenamefont {Yang}\ \emph {et~al.}(2019)\citenamefont {Yang}, \citenamefont {Gao}, \citenamefont {Wu}, \citenamefont {Yang}, \citenamefont {Sun}, \citenamefont {Zhang}, \citenamefont {Wang}, \citenamefont {Liu}, \citenamefont {Han}, \citenamefont {Yang},\ and\ \citenamefont {Li}}]{yangRecentAdvancesQuantum2019}%
  \BibitemOpen
  \bibfield  {author} {\bibinfo {author} {\bibfnamefont {Z.}~\bibnamefont {Yang}}, \bibinfo {author} {\bibfnamefont {M.}~\bibnamefont {Gao}}, \bibinfo {author} {\bibfnamefont {W.}~\bibnamefont {Wu}}, \bibinfo {author} {\bibfnamefont {X.}~\bibnamefont {Yang}}, \bibinfo {author} {\bibfnamefont {X.~W.}\ \bibnamefont {Sun}}, \bibinfo {author} {\bibfnamefont {J.}~\bibnamefont {Zhang}}, \bibinfo {author} {\bibfnamefont {H.-C.}\ \bibnamefont {Wang}}, \bibinfo {author} {\bibfnamefont {R.-S.}\ \bibnamefont {Liu}}, \bibinfo {author} {\bibfnamefont {C.-Y.}\ \bibnamefont {Han}}, \bibinfo {author} {\bibfnamefont {H.}~\bibnamefont {Yang}},\ and\ \bibinfo {author} {\bibfnamefont {W.}~\bibnamefont {Li}},\ }\bibfield  {title} {\bibinfo {title} {Recent advances in quantum dot-based light-emitting devices: {{Challenges}} and possible solutions},\ }\href {https://doi.org/10.1016/j.mattod.2018.09.002} {\bibfield  {journal} {\bibinfo  {journal} {Mater. Today}\ }\textbf {\bibinfo {volume} {24}},\ \bibinfo {pages} {69} (\bibinfo {year} {2019})}\BibitemShut {NoStop}%
\bibitem [{\citenamefont {Qu}\ \emph {et~al.}(2022)\citenamefont {Qu}, \citenamefont {Weis}, \citenamefont {Izquierdo}, \citenamefont {Mizrahi}, \citenamefont {Chu}, \citenamefont {Dabard}, \citenamefont {Gr{\'e}boval}, \citenamefont {Bossavit}, \citenamefont {Prado}, \citenamefont {P{\'e}ronne}, \citenamefont {Ithurria}, \citenamefont {Patriarche}, \citenamefont {Silly}, \citenamefont {Vincent}, \citenamefont {Boschetto},\ and\ \citenamefont {Lhuillier}}]{quElectroluminescenceNanocrystalsUm2022}%
  \BibitemOpen
  \bibfield  {author} {\bibinfo {author} {\bibfnamefont {J.}~\bibnamefont {Qu}}, \bibinfo {author} {\bibfnamefont {M.}~\bibnamefont {Weis}}, \bibinfo {author} {\bibfnamefont {E.}~\bibnamefont {Izquierdo}}, \bibinfo {author} {\bibfnamefont {S.~G.}\ \bibnamefont {Mizrahi}}, \bibinfo {author} {\bibfnamefont {A.}~\bibnamefont {Chu}}, \bibinfo {author} {\bibfnamefont {C.}~\bibnamefont {Dabard}}, \bibinfo {author} {\bibfnamefont {C.}~\bibnamefont {Gr{\'e}boval}}, \bibinfo {author} {\bibfnamefont {E.}~\bibnamefont {Bossavit}}, \bibinfo {author} {\bibfnamefont {Y.}~\bibnamefont {Prado}}, \bibinfo {author} {\bibfnamefont {E.}~\bibnamefont {P{\'e}ronne}}, \bibinfo {author} {\bibfnamefont {S.}~\bibnamefont {Ithurria}}, \bibinfo {author} {\bibfnamefont {G.}~\bibnamefont {Patriarche}}, \bibinfo {author} {\bibfnamefont {M.~G.}\ \bibnamefont {Silly}}, \bibinfo {author} {\bibfnamefont {G.}~\bibnamefont {Vincent}}, \bibinfo {author} {\bibfnamefont {D.}~\bibnamefont {Boschetto}},\ and\ \bibinfo {author} {\bibfnamefont {E.}~\bibnamefont {Lhuillier}},\ }\bibfield  {title} {\bibinfo {title} {Electroluminescence from nanocrystals above 2 {$M$}m},\ }\href {https://doi.org/10.1038/s41566-021-00902-y} {\bibfield  {journal} {\bibinfo  {journal} {Nat. Photonics}\ }\textbf {\bibinfo {volume} {16}},\ \bibinfo {pages} {38} (\bibinfo {year} {2022})}\BibitemShut {NoStop}%
\bibitem [{\citenamefont {Roh}\ \emph {et~al.}(2020)\citenamefont {Roh}, \citenamefont {Park}, \citenamefont {Lim},\ and\ \citenamefont {Klimov}}]{rohOpticallyPumpedColloidalquantumdot2020}%
  \BibitemOpen
  \bibfield  {author} {\bibinfo {author} {\bibfnamefont {J.}~\bibnamefont {Roh}}, \bibinfo {author} {\bibfnamefont {Y.-S.}\ \bibnamefont {Park}}, \bibinfo {author} {\bibfnamefont {J.}~\bibnamefont {Lim}},\ and\ \bibinfo {author} {\bibfnamefont {V.~I.}\ \bibnamefont {Klimov}},\ }\bibfield  {title} {\bibinfo {title} {Optically pumped colloidal-quantum-dot lasing in {{LED-like}} devices with an integrated optical cavity},\ }\href {https://doi.org/10.1038/s41467-019-14014-3} {\bibfield  {journal} {\bibinfo  {journal} {Nat. Commun.}\ }\textbf {\bibinfo {volume} {11}},\ \bibinfo {pages} {271} (\bibinfo {year} {2020})}\BibitemShut {NoStop}%
\bibitem [{\citenamefont {Jung}\ \emph {et~al.}(2021)\citenamefont {Jung}, \citenamefont {Ahn},\ and\ \citenamefont {Klimov}}]{jungProspectsChallengesColloidal2021}%
  \BibitemOpen
  \bibfield  {author} {\bibinfo {author} {\bibfnamefont {H.}~\bibnamefont {Jung}}, \bibinfo {author} {\bibfnamefont {N.}~\bibnamefont {Ahn}},\ and\ \bibinfo {author} {\bibfnamefont {V.~I.}\ \bibnamefont {Klimov}},\ }\bibfield  {title} {\bibinfo {title} {Prospects and challenges of colloidal quantum dot laser diodes},\ }\href {https://doi.org/10.1038/s41566-021-00827-6} {\bibfield  {journal} {\bibinfo  {journal} {Nat. Photonics}\ }\textbf {\bibinfo {volume} {15}},\ \bibinfo {pages} {643} (\bibinfo {year} {2021})}\BibitemShut {NoStop}%
\bibitem [{\citenamefont {Freeman}\ and\ \citenamefont {Willner}(2012)}]{freemanOpticalMolecularSensing2012}%
  \BibitemOpen
  \bibfield  {author} {\bibinfo {author} {\bibfnamefont {R.}~\bibnamefont {Freeman}}\ and\ \bibinfo {author} {\bibfnamefont {I.}~\bibnamefont {Willner}},\ }\bibfield  {title} {\bibinfo {title} {Optical molecular sensing with semiconductor quantum dots ({{QDs}})},\ }\href {https://doi.org/10.1039/C2CS15357B} {\bibfield  {journal} {\bibinfo  {journal} {Chem. Soc. Rev.}\ }\textbf {\bibinfo {volume} {41}},\ \bibinfo {pages} {4067} (\bibinfo {year} {2012})}\BibitemShut {NoStop}%
\bibitem [{\citenamefont {{Fery-Forgues}}(2013)}]{fery-forguesFluorescentOrganicNanocrystals2013a}%
  \BibitemOpen
  \bibfield  {author} {\bibinfo {author} {\bibfnamefont {S.}~\bibnamefont {{Fery-Forgues}}},\ }\bibfield  {title} {\bibinfo {title} {Fluorescent organic nanocrystals and non-doped nanoparticles for biological applications},\ }\href {https://doi.org/10.1039/C3NR02657D} {\bibfield  {journal} {\bibinfo  {journal} {Nanoscale}\ }\textbf {\bibinfo {volume} {5}},\ \bibinfo {pages} {8428} (\bibinfo {year} {2013})}\BibitemShut {NoStop}%
\bibitem [{\citenamefont {Zhu}\ \emph {et~al.}(2017)\citenamefont {Zhu}, \citenamefont {Song}, \citenamefont {Wang}, \citenamefont {Wan}, \citenamefont {Zhang}, \citenamefont {Ning},\ and\ \citenamefont {Yang}}]{zhuPhotoluminescenceMechanismGraphene2017}%
  \BibitemOpen
  \bibfield  {author} {\bibinfo {author} {\bibfnamefont {S.}~\bibnamefont {Zhu}}, \bibinfo {author} {\bibfnamefont {Y.}~\bibnamefont {Song}}, \bibinfo {author} {\bibfnamefont {J.}~\bibnamefont {Wang}}, \bibinfo {author} {\bibfnamefont {H.}~\bibnamefont {Wan}}, \bibinfo {author} {\bibfnamefont {Y.}~\bibnamefont {Zhang}}, \bibinfo {author} {\bibfnamefont {Y.}~\bibnamefont {Ning}},\ and\ \bibinfo {author} {\bibfnamefont {B.}~\bibnamefont {Yang}},\ }\bibfield  {title} {\bibinfo {title} {Photoluminescence mechanism in graphene quantum dots: {{Quantum}} confinement effect and surface/edge state},\ }\href {https://doi.org/10.1016/j.nantod.2016.12.006} {\bibfield  {journal} {\bibinfo  {journal} {Nano Today}\ }\textbf {\bibinfo {volume} {13}},\ \bibinfo {pages} {10} (\bibinfo {year} {2017})}\BibitemShut {NoStop}%
\bibitem [{\citenamefont {Dong}\ \emph {et~al.}(2018)\citenamefont {Dong}, \citenamefont {Qiao}, \citenamefont {Kim}, \citenamefont {Parobek}, \citenamefont {Rossi},\ and\ \citenamefont {Son}}]{dongPreciseControlQuantum2018a}%
  \BibitemOpen
  \bibfield  {author} {\bibinfo {author} {\bibfnamefont {Y.}~\bibnamefont {Dong}}, \bibinfo {author} {\bibfnamefont {T.}~\bibnamefont {Qiao}}, \bibinfo {author} {\bibfnamefont {D.}~\bibnamefont {Kim}}, \bibinfo {author} {\bibfnamefont {D.}~\bibnamefont {Parobek}}, \bibinfo {author} {\bibfnamefont {D.}~\bibnamefont {Rossi}},\ and\ \bibinfo {author} {\bibfnamefont {D.~H.}\ \bibnamefont {Son}},\ }\bibfield  {title} {\bibinfo {title} {Precise {{Control}} of {{Quantum Confinement}} in {{Cesium Lead Halide Perovskite Quantum Dots}} via {{Thermodynamic Equilibrium}}},\ }\href {https://doi.org/10.1021/ACS.nanolett.8b00861} {\bibfield  {journal} {\bibinfo  {journal} {Nano Lett.}\ }\textbf {\bibinfo {volume} {18}},\ \bibinfo {pages} {3716} (\bibinfo {year} {2018})}\BibitemShut {NoStop}%
\bibitem [{\citenamefont {Zhang}\ and\ \citenamefont {Lin}(2012)}]{zhangDefectrelatedLuminescentMaterials2012}%
  \BibitemOpen
  \bibfield  {author} {\bibinfo {author} {\bibfnamefont {C.}~\bibnamefont {Zhang}}\ and\ \bibinfo {author} {\bibfnamefont {J.}~\bibnamefont {Lin}},\ }\bibfield  {title} {\bibinfo {title} {Defect-related luminescent materials: Synthesis, emission properties and applications},\ }\href {https://doi.org/10.1039/C2CS35215J} {\bibfield  {journal} {\bibinfo  {journal} {Chem. Soc. Rev.}\ }\textbf {\bibinfo {volume} {41}},\ \bibinfo {pages} {7938} (\bibinfo {year} {2012})}\BibitemShut {NoStop}%
\bibitem [{\citenamefont {Zhou}\ \emph {et~al.}(2022)\citenamefont {Zhou}, \citenamefont {Xie}, \citenamefont {Wang}, \citenamefont {Liang}, \citenamefont {Ou},\ and\ \citenamefont {Zhang}}]{zhouHighEfficiencyFastRadiativeBlueEmitting2022}%
  \BibitemOpen
  \bibfield  {author} {\bibinfo {author} {\bibfnamefont {A.}~\bibnamefont {Zhou}}, \bibinfo {author} {\bibfnamefont {Y.}~\bibnamefont {Xie}}, \bibinfo {author} {\bibfnamefont {F.}~\bibnamefont {Wang}}, \bibinfo {author} {\bibfnamefont {R.}~\bibnamefont {Liang}}, \bibinfo {author} {\bibfnamefont {Q.}~\bibnamefont {Ou}},\ and\ \bibinfo {author} {\bibfnamefont {S.}~\bibnamefont {Zhang}},\ }\bibfield  {title} {\bibinfo {title} {High-{{Efficiency Fast-Radiative Blue-Emitting Perovskite Nanoplatelets}} and {{Their Formation Mechanisms}}},\ }\href {https://doi.org/10.1021/ACS.jpclett.2c01041} {\bibfield  {journal} {\bibinfo  {journal} {J. Phys. Chem. Lett.}\ ,\ \bibinfo {pages} {4634}} (\bibinfo {year} {2022})}\BibitemShut {NoStop}%
\bibitem [{\citenamefont {Benin}\ \emph {et~al.}(2018)\citenamefont {Benin}, \citenamefont {Dirin}, \citenamefont {Morad}, \citenamefont {W{\"o}rle}, \citenamefont {Yakunin}, \citenamefont {Rain{\`o}}, \citenamefont {Nazarenko}, \citenamefont {Fischer}, \citenamefont {Infante},\ and\ \citenamefont {Kovalenko}}]{beninHighlyEmissiveSelfTrapped2018}%
  \BibitemOpen
  \bibfield  {author} {\bibinfo {author} {\bibfnamefont {B.~M.}\ \bibnamefont {Benin}}, \bibinfo {author} {\bibfnamefont {D.~N.}\ \bibnamefont {Dirin}}, \bibinfo {author} {\bibfnamefont {V.}~\bibnamefont {Morad}}, \bibinfo {author} {\bibfnamefont {M.}~\bibnamefont {W{\"o}rle}}, \bibinfo {author} {\bibfnamefont {S.}~\bibnamefont {Yakunin}}, \bibinfo {author} {\bibfnamefont {G.}~\bibnamefont {Rain{\`o}}}, \bibinfo {author} {\bibfnamefont {O.}~\bibnamefont {Nazarenko}}, \bibinfo {author} {\bibfnamefont {M.}~\bibnamefont {Fischer}}, \bibinfo {author} {\bibfnamefont {I.}~\bibnamefont {Infante}},\ and\ \bibinfo {author} {\bibfnamefont {M.~V.}\ \bibnamefont {Kovalenko}},\ }\bibfield  {title} {\bibinfo {title} {Highly {{Emissive Self-Trapped Excitons}} in {{Fully Inorganic Zero-Dimensional Tin Halides}}},\ }\href {https://doi.org/10.1002/anie.201806452} {\bibfield  {journal} {\bibinfo  {journal} {Angew. Chem. Int. Ed.}\ }\textbf {\bibinfo {volume} {57}},\ \bibinfo {pages} {11329} (\bibinfo {year} {2018})}\BibitemShut {NoStop}%
\bibitem [{\citenamefont {Jing}\ \emph {et~al.}(2021)\citenamefont {Jing}, \citenamefont {Liu}, \citenamefont {Li},\ and\ \citenamefont {Xia}}]{jingPhotoluminescenceSingletTriplet2021}%
  \BibitemOpen
  \bibfield  {author} {\bibinfo {author} {\bibfnamefont {Y.}~\bibnamefont {Jing}}, \bibinfo {author} {\bibfnamefont {Y.}~\bibnamefont {Liu}}, \bibinfo {author} {\bibfnamefont {M.}~\bibnamefont {Li}},\ and\ \bibinfo {author} {\bibfnamefont {Z.}~\bibnamefont {Xia}},\ }\bibfield  {title} {\bibinfo {title} {Photoluminescence of {{Singlet}}/{{Triplet Self-Trapped Excitons}} in {{Sb3}}+-{{Based Metal Halides}}},\ }\href {https://doi.org/10.1002/adom.202002213} {\bibfield  {journal} {\bibinfo  {journal} {Adv. Opt. Mater.}\ }\textbf {\bibinfo {volume} {9}},\ \bibinfo {pages} {2002213} (\bibinfo {year} {2021})}\BibitemShut {NoStop}%
\bibitem [{\citenamefont {Luo}\ \emph {et~al.}(2001)\citenamefont {Luo}, \citenamefont {Xie}, \citenamefont {Lam}, \citenamefont {Cheng}, \citenamefont {Chen}, \citenamefont {Qiu}, \citenamefont {Kwok}, \citenamefont {Zhan}, \citenamefont {Liu}, \citenamefont {Zhu},\ and\ \citenamefont {Tang}}]{luoAggregationinducedEmission1methyl12001}%
  \BibitemOpen
  \bibfield  {author} {\bibinfo {author} {\bibfnamefont {J.}~\bibnamefont {Luo}}, \bibinfo {author} {\bibfnamefont {Z.}~\bibnamefont {Xie}}, \bibinfo {author} {\bibfnamefont {J.~W.~Y.}\ \bibnamefont {Lam}}, \bibinfo {author} {\bibfnamefont {L.}~\bibnamefont {Cheng}}, \bibinfo {author} {\bibfnamefont {H.}~\bibnamefont {Chen}}, \bibinfo {author} {\bibfnamefont {C.}~\bibnamefont {Qiu}}, \bibinfo {author} {\bibfnamefont {H.~S.}\ \bibnamefont {Kwok}}, \bibinfo {author} {\bibfnamefont {X.}~\bibnamefont {Zhan}}, \bibinfo {author} {\bibfnamefont {Y.}~\bibnamefont {Liu}}, \bibinfo {author} {\bibfnamefont {D.}~\bibnamefont {Zhu}},\ and\ \bibinfo {author} {\bibfnamefont {B.~Z.}\ \bibnamefont {Tang}},\ }\bibfield  {title} {\bibinfo {title} {Aggregation-induced emission of 1-methyl-1,2,3,4,5-pentaphenylsilole},\ }\href {https://doi.org/10.1039/B105159H} {\bibfield  {journal} {\bibinfo  {journal} {Chem. Commun.}\ ,\ \bibinfo {pages} {1740}} (\bibinfo {year} {2001})}\BibitemShut {NoStop}%
\bibitem [{\citenamefont {Zhang}\ \emph {et~al.}(2020)\citenamefont {Zhang}, \citenamefont {Zhao}, \citenamefont {Turley}, \citenamefont {Wang}, \citenamefont {McGonigal}, \citenamefont {Tu}, \citenamefont {Li}, \citenamefont {Wang}, \citenamefont {Kwok}, \citenamefont {Lam},\ and\ \citenamefont {Tang}}]{zhangAggregateScienceStructures2020}%
  \BibitemOpen
  \bibfield  {author} {\bibinfo {author} {\bibfnamefont {H.}~\bibnamefont {Zhang}}, \bibinfo {author} {\bibfnamefont {Z.}~\bibnamefont {Zhao}}, \bibinfo {author} {\bibfnamefont {A.~T.}\ \bibnamefont {Turley}}, \bibinfo {author} {\bibfnamefont {L.}~\bibnamefont {Wang}}, \bibinfo {author} {\bibfnamefont {P.~R.}\ \bibnamefont {McGonigal}}, \bibinfo {author} {\bibfnamefont {Y.}~\bibnamefont {Tu}}, \bibinfo {author} {\bibfnamefont {Y.}~\bibnamefont {Li}}, \bibinfo {author} {\bibfnamefont {Z.}~\bibnamefont {Wang}}, \bibinfo {author} {\bibfnamefont {R.~T.~K.}\ \bibnamefont {Kwok}}, \bibinfo {author} {\bibfnamefont {J.~W.~Y.}\ \bibnamefont {Lam}},\ and\ \bibinfo {author} {\bibfnamefont {B.~Z.}\ \bibnamefont {Tang}},\ }\bibfield  {title} {\bibinfo {title} {Aggregate {{Science}}: {{From Structures}} to {{Properties}}},\ }\href {https://doi.org/10.1002/adma.202001457} {\bibfield  {journal} {\bibinfo  {journal} {Adv. Mater.}\ }\textbf {\bibinfo {volume} {32}},\ \bibinfo {pages} {2001457} (\bibinfo {year} {2020})}\BibitemShut {NoStop}%
\bibitem [{\citenamefont {Pringle}\ \emph {et~al.}(2020)\citenamefont {Pringle}, \citenamefont {Hunter}, \citenamefont {Brumberg}, \citenamefont {Anderson}, \citenamefont {Fagan}, \citenamefont {Thomas}, \citenamefont {Petersen}, \citenamefont {Sefannaser}, \citenamefont {Han}, \citenamefont {Brown}, \citenamefont {Kilin}, \citenamefont {Schaller}, \citenamefont {Kortshagen}, \citenamefont {Boudjouk},\ and\ \citenamefont {Hobbie}}]{pringleBrightSiliconNanocrystals2020a}%
  \BibitemOpen
  \bibfield  {author} {\bibinfo {author} {\bibfnamefont {T.~A.}\ \bibnamefont {Pringle}}, \bibinfo {author} {\bibfnamefont {K.~I.}\ \bibnamefont {Hunter}}, \bibinfo {author} {\bibfnamefont {A.}~\bibnamefont {Brumberg}}, \bibinfo {author} {\bibfnamefont {K.~J.}\ \bibnamefont {Anderson}}, \bibinfo {author} {\bibfnamefont {J.~A.}\ \bibnamefont {Fagan}}, \bibinfo {author} {\bibfnamefont {S.~A.}\ \bibnamefont {Thomas}}, \bibinfo {author} {\bibfnamefont {R.~J.}\ \bibnamefont {Petersen}}, \bibinfo {author} {\bibfnamefont {M.}~\bibnamefont {Sefannaser}}, \bibinfo {author} {\bibfnamefont {Y.}~\bibnamefont {Han}}, \bibinfo {author} {\bibfnamefont {S.~L.}\ \bibnamefont {Brown}}, \bibinfo {author} {\bibfnamefont {D.~S.}\ \bibnamefont {Kilin}}, \bibinfo {author} {\bibfnamefont {R.~D.}\ \bibnamefont {Schaller}}, \bibinfo {author} {\bibfnamefont {U.~R.}\ \bibnamefont {Kortshagen}}, \bibinfo {author} {\bibfnamefont {P.~R.}\ \bibnamefont {Boudjouk}},\ and\ \bibinfo {author} {\bibfnamefont {E.~K.}\ \bibnamefont {Hobbie}},\ }\bibfield  {title} {\bibinfo {title} {Bright {{Silicon Nanocrystals}} from a {{Liquid Precursor}}: {{Quasi-Direct Recombination}} with {{High Quantum Yield}}},\ }\href {https://doi.org/10/ghdnmp} {\bibfield  {journal} {\bibinfo  {journal} {ACS Nano}\ }\textbf {\bibinfo {volume} {14}},\ \bibinfo {pages} {3858} (\bibinfo {year} {2020})}\BibitemShut {NoStop}%
\bibitem [{\citenamefont {Du}\ \emph {et~al.}(2017)\citenamefont {Du}, \citenamefont {Meng}, \citenamefont {Wang}, \citenamefont {Yan},\ and\ \citenamefont {Mitzi}}]{duBandgapEngineeringLeadFree2017}%
  \BibitemOpen
  \bibfield  {author} {\bibinfo {author} {\bibfnamefont {K.-z.}\ \bibnamefont {Du}}, \bibinfo {author} {\bibfnamefont {W.}~\bibnamefont {Meng}}, \bibinfo {author} {\bibfnamefont {X.}~\bibnamefont {Wang}}, \bibinfo {author} {\bibfnamefont {Y.}~\bibnamefont {Yan}},\ and\ \bibinfo {author} {\bibfnamefont {D.~B.}\ \bibnamefont {Mitzi}},\ }\bibfield  {title} {\bibinfo {title} {Bandgap {{Engineering}} of {{Lead-Free Double Perovskite Cs2AgBiBr6}} through {{Trivalent Metal Alloying}}},\ }\href {https://doi.org/10.1002/anie.201703970} {\bibfield  {journal} {\bibinfo  {journal} {Angew. Chem. Int. Ed.}\ }\textbf {\bibinfo {volume} {56}},\ \bibinfo {pages} {8158} (\bibinfo {year} {2017})}\BibitemShut {NoStop}%
\bibitem [{\citenamefont {Wang}\ \emph {et~al.}(2022)\citenamefont {Wang}, \citenamefont {Ou},\ and\ \citenamefont {Zhang}}]{wangEnhancementIntrinsicOptical2022}%
  \BibitemOpen
  \bibfield  {author} {\bibinfo {author} {\bibfnamefont {F.}~\bibnamefont {Wang}}, \bibinfo {author} {\bibfnamefont {Q.}~\bibnamefont {Ou}},\ and\ \bibinfo {author} {\bibfnamefont {S.}~\bibnamefont {Zhang}},\ }\bibfield  {title} {\bibinfo {title} {Enhancement of intrinsic optical transitions in silicon nanocrystals by state localization},\ }\href {https://doi.org/10.1103/PhysRevB.106.155425} {\bibfield  {journal} {\bibinfo  {journal} {Phys. Rev. B}\ }\textbf {\bibinfo {volume} {106}},\ \bibinfo {pages} {155425} (\bibinfo {year} {2022})}\BibitemShut {NoStop}%
\bibitem [{\citenamefont {Dey}\ \emph {et~al.}(2020)\citenamefont {Dey}, \citenamefont {Richter}, \citenamefont {Debnath}, \citenamefont {Huang}, \citenamefont {Polavarapu},\ and\ \citenamefont {Feldmann}}]{deyTransferDirectIndirect2020}%
  \BibitemOpen
  \bibfield  {author} {\bibinfo {author} {\bibfnamefont {A.}~\bibnamefont {Dey}}, \bibinfo {author} {\bibfnamefont {A.~F.}\ \bibnamefont {Richter}}, \bibinfo {author} {\bibfnamefont {T.}~\bibnamefont {Debnath}}, \bibinfo {author} {\bibfnamefont {H.}~\bibnamefont {Huang}}, \bibinfo {author} {\bibfnamefont {L.}~\bibnamefont {Polavarapu}},\ and\ \bibinfo {author} {\bibfnamefont {J.}~\bibnamefont {Feldmann}},\ }\bibfield  {title} {\bibinfo {title} {Transfer of {{Direct}} to {{Indirect Bound Excitons}} by {{Electron Intervalley Scattering}} in {{Cs2AgBiBr6 Double Perovskite Nanocrystals}}},\ }\href {https://doi.org/10.1021/ACSnano.0c00997} {\bibfield  {journal} {\bibinfo  {journal} {ACS Nano}\ }\textbf {\bibinfo {volume} {14}},\ \bibinfo {pages} {5855} (\bibinfo {year} {2020})}\BibitemShut {NoStop}%
\bibitem [{\citenamefont {Luo}\ \emph {et~al.}(2018)\citenamefont {Luo}, \citenamefont {Wang}, \citenamefont {Li}, \citenamefont {Liu}, \citenamefont {Guo}, \citenamefont {Niu}, \citenamefont {Yao}, \citenamefont {Fu}, \citenamefont {Gao}, \citenamefont {Dong}, \citenamefont {Zhao}, \citenamefont {Leng}, \citenamefont {Ma}, \citenamefont {Liang}, \citenamefont {Wang}, \citenamefont {Jin}, \citenamefont {Han}, \citenamefont {Zhang}, \citenamefont {Etheridge}, \citenamefont {Wang}, \citenamefont {Yan}, \citenamefont {Sargent},\ and\ \citenamefont {Tang}}]{luoEfficientStableEmission2018}%
  \BibitemOpen
  \bibfield  {author} {\bibinfo {author} {\bibfnamefont {J.}~\bibnamefont {Luo}}, \bibinfo {author} {\bibfnamefont {X.}~\bibnamefont {Wang}}, \bibinfo {author} {\bibfnamefont {S.}~\bibnamefont {Li}}, \bibinfo {author} {\bibfnamefont {J.}~\bibnamefont {Liu}}, \bibinfo {author} {\bibfnamefont {Y.}~\bibnamefont {Guo}}, \bibinfo {author} {\bibfnamefont {G.}~\bibnamefont {Niu}}, \bibinfo {author} {\bibfnamefont {L.}~\bibnamefont {Yao}}, \bibinfo {author} {\bibfnamefont {Y.}~\bibnamefont {Fu}}, \bibinfo {author} {\bibfnamefont {L.}~\bibnamefont {Gao}}, \bibinfo {author} {\bibfnamefont {Q.}~\bibnamefont {Dong}}, \bibinfo {author} {\bibfnamefont {C.}~\bibnamefont {Zhao}}, \bibinfo {author} {\bibfnamefont {M.}~\bibnamefont {Leng}}, \bibinfo {author} {\bibfnamefont {F.}~\bibnamefont {Ma}}, \bibinfo {author} {\bibfnamefont {W.}~\bibnamefont {Liang}}, \bibinfo {author} {\bibfnamefont {L.}~\bibnamefont {Wang}}, \bibinfo {author} {\bibfnamefont {S.}~\bibnamefont {Jin}}, \bibinfo {author} {\bibfnamefont {J.}~\bibnamefont {Han}}, \bibinfo {author} {\bibfnamefont {L.}~\bibnamefont {Zhang}}, \bibinfo {author} {\bibfnamefont {J.}~\bibnamefont {Etheridge}}, \bibinfo {author} {\bibfnamefont {J.}~\bibnamefont {Wang}}, \bibinfo {author} {\bibfnamefont {Y.}~\bibnamefont {Yan}}, \bibinfo {author} {\bibfnamefont {E.~H.}\ \bibnamefont {Sargent}},\ and\ \bibinfo {author} {\bibfnamefont {J.}~\bibnamefont {Tang}},\ }\bibfield  {title} {\bibinfo {title} {Efficient and stable emission of warm-white light from lead-free halide double perovskites},\ }\href {https://doi.org/10.1038/s41586-018-0691-0} {\bibfield  {journal} {\bibinfo  {journal} {Nature}\ }\textbf {\bibinfo {volume} {563}},\ \bibinfo {pages} {541} (\bibinfo {year} {2018})}\BibitemShut {NoStop}%
\bibitem [{\citenamefont {Han}\ \emph {et~al.}(2019)\citenamefont {Han}, \citenamefont {Mao}, \citenamefont {Yang}, \citenamefont {Zhang}, \citenamefont {Yang}, \citenamefont {Wei}, \citenamefont {Deng},\ and\ \citenamefont {Han}}]{hanLeadFreeSodiumIndium2019}%
  \BibitemOpen
  \bibfield  {author} {\bibinfo {author} {\bibfnamefont {P.}~\bibnamefont {Han}}, \bibinfo {author} {\bibfnamefont {X.}~\bibnamefont {Mao}}, \bibinfo {author} {\bibfnamefont {S.}~\bibnamefont {Yang}}, \bibinfo {author} {\bibfnamefont {F.}~\bibnamefont {Zhang}}, \bibinfo {author} {\bibfnamefont {B.}~\bibnamefont {Yang}}, \bibinfo {author} {\bibfnamefont {D.}~\bibnamefont {Wei}}, \bibinfo {author} {\bibfnamefont {W.}~\bibnamefont {Deng}},\ and\ \bibinfo {author} {\bibfnamefont {K.}~\bibnamefont {Han}},\ }\bibfield  {title} {\bibinfo {title} {Lead-{{Free Sodium}}\textendash{{Indium Double Perovskite Nanocrystals}} through {{Doping Silver Cations}} for {{Bright Yellow Emission}}},\ }\href {https://doi.org/10.1002/anie.201909525} {\bibfield  {journal} {\bibinfo  {journal} {Angew. Chem. Int. Ed.}\ }\textbf {\bibinfo {volume} {58}},\ \bibinfo {pages} {17231} (\bibinfo {year} {2019})}\BibitemShut {NoStop}%
\bibitem [{Elk()}]{ElkCode}%
  \BibitemOpen
  \href@noop {} {\bibinfo {title} {The {{Elk Code}}}},\ \bibinfo {howpublished} {https://elk.sourceforge.io/}\BibitemShut {NoStop}%
\bibitem [{\citenamefont {Perdew}\ \emph {et~al.}(1996{\natexlab{a}})\citenamefont {Perdew}, \citenamefont {Burke},\ and\ \citenamefont {Ernzerhof}}]{perdewGeneralizedGradientApproximation1996}%
  \BibitemOpen
  \bibfield  {author} {\bibinfo {author} {\bibfnamefont {J.~P.}\ \bibnamefont {Perdew}}, \bibinfo {author} {\bibfnamefont {K.}~\bibnamefont {Burke}},\ and\ \bibinfo {author} {\bibfnamefont {M.}~\bibnamefont {Ernzerhof}},\ }\bibfield  {title} {\bibinfo {title} {Generalized {{Gradient Approximation Made Simple}}},\ }\href {https://doi.org/10/bppfwt} {\bibfield  {journal} {\bibinfo  {journal} {Phys. Rev. Lett.}\ }\textbf {\bibinfo {volume} {77}},\ \bibinfo {pages} {3865} (\bibinfo {year} {1996}{\natexlab{a}})}\BibitemShut {NoStop}%
\bibitem [{\citenamefont {Furness}\ \emph {et~al.}(2020)\citenamefont {Furness}, \citenamefont {Kaplan}, \citenamefont {Ning}, \citenamefont {Perdew},\ and\ \citenamefont {Sun}}]{furnessAccurateNumericallyEfficient2020}%
  \BibitemOpen
  \bibfield  {author} {\bibinfo {author} {\bibfnamefont {J.~W.}\ \bibnamefont {Furness}}, \bibinfo {author} {\bibfnamefont {A.~D.}\ \bibnamefont {Kaplan}}, \bibinfo {author} {\bibfnamefont {J.}~\bibnamefont {Ning}}, \bibinfo {author} {\bibfnamefont {J.~P.}\ \bibnamefont {Perdew}},\ and\ \bibinfo {author} {\bibfnamefont {J.}~\bibnamefont {Sun}},\ }\bibfield  {title} {\bibinfo {title} {Accurate and {{Numerically Efficient}} r {\textsuperscript{2}} {{SCAN Meta-Generalized Gradient Approximation}}},\ }\href {https://doi.org/10.1021/ACS.jpclett.0c02405} {\bibfield  {journal} {\bibinfo  {journal} {J. Phys. Chem. Lett.}\ }\textbf {\bibinfo {volume} {11}},\ \bibinfo {pages} {8208} (\bibinfo {year} {2020})}\BibitemShut {NoStop}%
\bibitem [{\citenamefont {K{\"u}hne}\ \emph {et~al.}(2020)\citenamefont {K{\"u}hne}, \citenamefont {Iannuzzi}, \citenamefont {Del~Ben}, \citenamefont {Rybkin}, \citenamefont {Seewald}, \citenamefont {Stein}, \citenamefont {Laino}, \citenamefont {Khaliullin}, \citenamefont {Sch{\"u}tt}, \citenamefont {Schiffmann}, \citenamefont {Golze}, \citenamefont {Wilhelm}, \citenamefont {Chulkov}, \citenamefont {{Bani-Hashemian}}, \citenamefont {Weber}, \citenamefont {Bor{\v s}tnik}, \citenamefont {Taillefumier}, \citenamefont {Jakobovits}, \citenamefont {Lazzaro}, \citenamefont {Pabst}, \citenamefont {M{\"u}ller}, \citenamefont {Schade}, \citenamefont {Guidon}, \citenamefont {Andermatt}, \citenamefont {Holmberg}, \citenamefont {Schenter}, \citenamefont {Hehn}, \citenamefont {Bussy}, \citenamefont {Belleflamme}, \citenamefont {Tabacchi}, \citenamefont {Gl{\"o}{\ss}}, \citenamefont {Lass}, \citenamefont {Bethune}, \citenamefont {Mundy}, \citenamefont {Plessl}, \citenamefont {Watkins}, \citenamefont {VandeVondele}, \citenamefont {Krack},\ and\ \citenamefont {Hutter}}]{kuhneCP2KElectronicStructure2020}%
  \BibitemOpen
  \bibfield  {author} {\bibinfo {author} {\bibfnamefont {T.~D.}\ \bibnamefont {K{\"u}hne}}, \bibinfo {author} {\bibfnamefont {M.}~\bibnamefont {Iannuzzi}}, \bibinfo {author} {\bibfnamefont {M.}~\bibnamefont {Del~Ben}}, \bibinfo {author} {\bibfnamefont {V.~V.}\ \bibnamefont {Rybkin}}, \bibinfo {author} {\bibfnamefont {P.}~\bibnamefont {Seewald}}, \bibinfo {author} {\bibfnamefont {F.}~\bibnamefont {Stein}}, \bibinfo {author} {\bibfnamefont {T.}~\bibnamefont {Laino}}, \bibinfo {author} {\bibfnamefont {R.~Z.}\ \bibnamefont {Khaliullin}}, \bibinfo {author} {\bibfnamefont {O.}~\bibnamefont {Sch{\"u}tt}}, \bibinfo {author} {\bibfnamefont {F.}~\bibnamefont {Schiffmann}}, \bibinfo {author} {\bibfnamefont {D.}~\bibnamefont {Golze}}, \bibinfo {author} {\bibfnamefont {J.}~\bibnamefont {Wilhelm}}, \bibinfo {author} {\bibfnamefont {S.}~\bibnamefont {Chulkov}}, \bibinfo {author} {\bibfnamefont {M.~H.}\ \bibnamefont {{Bani-Hashemian}}}, \bibinfo {author} {\bibfnamefont {V.}~\bibnamefont {Weber}}, \bibinfo {author} {\bibfnamefont {U.}~\bibnamefont {Bor{\v s}tnik}}, \bibinfo {author} {\bibfnamefont {M.}~\bibnamefont {Taillefumier}}, \bibinfo {author} {\bibfnamefont {A.~S.}\ \bibnamefont {Jakobovits}}, \bibinfo {author} {\bibfnamefont {A.}~\bibnamefont {Lazzaro}}, \bibinfo {author} {\bibfnamefont {H.}~\bibnamefont {Pabst}}, \bibinfo {author} {\bibfnamefont {T.}~\bibnamefont {M{\"u}ller}}, \bibinfo {author} {\bibfnamefont {R.}~\bibnamefont {Schade}}, \bibinfo {author} {\bibfnamefont {M.}~\bibnamefont {Guidon}}, \bibinfo {author} {\bibfnamefont {S.}~\bibnamefont {Andermatt}}, \bibinfo {author} {\bibfnamefont {N.}~\bibnamefont {Holmberg}}, \bibinfo {author} {\bibfnamefont {G.~K.}\ \bibnamefont {Schenter}}, \bibinfo {author} {\bibfnamefont {A.}~\bibnamefont {Hehn}}, \bibinfo {author} {\bibfnamefont {A.}~\bibnamefont {Bussy}}, \bibinfo {author} {\bibfnamefont {F.}~\bibnamefont {Belleflamme}}, \bibinfo {author} {\bibfnamefont {G.}~\bibnamefont {Tabacchi}}, \bibinfo {author} {\bibfnamefont {A.}~\bibnamefont {Gl{\"o}{\ss}}}, \bibinfo {author} {\bibfnamefont {M.}~\bibnamefont {Lass}}, \bibinfo {author} {\bibfnamefont {I.}~\bibnamefont {Bethune}}, \bibinfo {author} {\bibfnamefont {C.~J.}\ \bibnamefont {Mundy}}, \bibinfo {author} {\bibfnamefont {C.}~\bibnamefont {Plessl}}, \bibinfo {author} {\bibfnamefont {M.}~\bibnamefont {Watkins}}, \bibinfo {author} {\bibfnamefont {J.}~\bibnamefont {VandeVondele}}, \bibinfo {author} {\bibfnamefont {M.}~\bibnamefont {Krack}},\ and\ \bibinfo {author} {\bibfnamefont {J.}~\bibnamefont {Hutter}},\ }\bibfield  {title} {\bibinfo {title} {{{CP2K}}: {{An}} electronic structure and molecular dynamics software package - {{Quickstep}}: {{Efficient}} and accurate electronic structure calculations},\ }\href {https://doi.org/10/gm783j} {\bibfield  {journal} {\bibinfo  {journal} {J. Chem. Phys.}\ }\textbf {\bibinfo {volume} {152}},\ \bibinfo {pages} {194103} (\bibinfo {year} {2020})}\BibitemShut {NoStop}%
\bibitem [{\citenamefont {VandeVondele}\ \emph {et~al.}(2005)\citenamefont {VandeVondele}, \citenamefont {Krack}, \citenamefont {Mohamed}, \citenamefont {Parrinello}, \citenamefont {Chassaing},\ and\ \citenamefont {Hutter}}]{vandevondeleQuickstepFastAccurate2005}%
  \BibitemOpen
  \bibfield  {author} {\bibinfo {author} {\bibfnamefont {J.}~\bibnamefont {VandeVondele}}, \bibinfo {author} {\bibfnamefont {M.}~\bibnamefont {Krack}}, \bibinfo {author} {\bibfnamefont {F.}~\bibnamefont {Mohamed}}, \bibinfo {author} {\bibfnamefont {M.}~\bibnamefont {Parrinello}}, \bibinfo {author} {\bibfnamefont {T.}~\bibnamefont {Chassaing}},\ and\ \bibinfo {author} {\bibfnamefont {J.}~\bibnamefont {Hutter}},\ }\bibfield  {title} {\bibinfo {title} {Quickstep: {{Fast}} and accurate density functional calculations using a mixed {{Gaussian}} and plane waves approach},\ }\href {https://doi.org/10.1016/j.cpc.2004.12.014} {\bibfield  {journal} {\bibinfo  {journal} {Comput. Phys. Comm.}\ }\textbf {\bibinfo {volume} {167}},\ \bibinfo {pages} {103} (\bibinfo {year} {2005})}\BibitemShut {NoStop}%
\bibitem [{\citenamefont {VandeVondele}\ and\ \citenamefont {Hutter}(2007)}]{vandevondeleGaussianBasisSets2007a}%
  \BibitemOpen
  \bibfield  {author} {\bibinfo {author} {\bibfnamefont {J.}~\bibnamefont {VandeVondele}}\ and\ \bibinfo {author} {\bibfnamefont {J.}~\bibnamefont {Hutter}},\ }\bibfield  {title} {\bibinfo {title} {Gaussian basis sets for accurate calculations on molecular systems in gas and condensed phases},\ }\href {https://doi.org/10/b3pxwr} {\bibfield  {journal} {\bibinfo  {journal} {J. Chem. Phys.}\ }\textbf {\bibinfo {volume} {127}},\ \bibinfo {pages} {114105} (\bibinfo {year} {2007})}\BibitemShut {NoStop}%
\bibitem [{\citenamefont {Goedecker}\ \emph {et~al.}(1996)\citenamefont {Goedecker}, \citenamefont {Teter},\ and\ \citenamefont {Hutter}}]{goedeckerSeparableDualspaceGaussian1996}%
  \BibitemOpen
  \bibfield  {author} {\bibinfo {author} {\bibfnamefont {S.}~\bibnamefont {Goedecker}}, \bibinfo {author} {\bibfnamefont {M.}~\bibnamefont {Teter}},\ and\ \bibinfo {author} {\bibfnamefont {J.}~\bibnamefont {Hutter}},\ }\bibfield  {title} {\bibinfo {title} {Separable dual-space {{Gaussian}} pseudopotentials},\ }\href {https://doi.org/10.1103/PhysRevB.54.1703} {\bibfield  {journal} {\bibinfo  {journal} {Phys. Rev. B}\ }\textbf {\bibinfo {volume} {54}},\ \bibinfo {pages} {1703} (\bibinfo {year} {1996})}\BibitemShut {NoStop}%
\bibitem [{\citenamefont {Krack}(2005)}]{krackPseudopotentialsKrOptimized2005}%
  \BibitemOpen
  \bibfield  {author} {\bibinfo {author} {\bibfnamefont {M.}~\bibnamefont {Krack}},\ }\bibfield  {title} {\bibinfo {title} {Pseudopotentials for {{H}} to {{Kr}} optimized for gradient-corrected exchange-correlation functionals},\ }\href {https://doi.org/10/b68xdh} {\bibfield  {journal} {\bibinfo  {journal} {Theor. Chem. Acc.}\ }\textbf {\bibinfo {volume} {114}},\ \bibinfo {pages} {145} (\bibinfo {year} {2005})}\BibitemShut {NoStop}%
\bibitem [{\citenamefont {Grimme}\ \emph {et~al.}(2010)\citenamefont {Grimme}, \citenamefont {Antony}, \citenamefont {Ehrlich},\ and\ \citenamefont {Krieg}}]{grimmeConsistentAccurateInitio2010b}%
  \BibitemOpen
  \bibfield  {author} {\bibinfo {author} {\bibfnamefont {S.}~\bibnamefont {Grimme}}, \bibinfo {author} {\bibfnamefont {J.}~\bibnamefont {Antony}}, \bibinfo {author} {\bibfnamefont {S.}~\bibnamefont {Ehrlich}},\ and\ \bibinfo {author} {\bibfnamefont {H.}~\bibnamefont {Krieg}},\ }\bibfield  {title} {\bibinfo {title} {A consistent and accurate ab initio parametrization of density functional dispersion correction ({{DFT-D}}) for the 94 elements {{H-Pu}}},\ }\href {https://doi.org/10.1063/1.3382344} {\bibfield  {journal} {\bibinfo  {journal} {J. Chem. Phys.}\ }\textbf {\bibinfo {volume} {132}},\ \bibinfo {pages} {154104} (\bibinfo {year} {2010})}\BibitemShut {NoStop}%
\bibitem [{\citenamefont {Perdew}\ \emph {et~al.}(1996{\natexlab{b}})\citenamefont {Perdew}, \citenamefont {Burke},\ and\ \citenamefont {Ernzerhof}}]{perdewGeneralizedGradientApproximation1996c}%
  \BibitemOpen
  \bibfield  {author} {\bibinfo {author} {\bibfnamefont {J.~P.}\ \bibnamefont {Perdew}}, \bibinfo {author} {\bibfnamefont {K.}~\bibnamefont {Burke}},\ and\ \bibinfo {author} {\bibfnamefont {M.}~\bibnamefont {Ernzerhof}},\ }\bibfield  {title} {\bibinfo {title} {Generalized gradient approximation made simple},\ }\href {https://doi.org/10.1103/PhysRevLett.77.3865} {\bibfield  {journal} {\bibinfo  {journal} {Physical Review Letters}\ }\textbf {\bibinfo {volume} {77}},\ \bibinfo {pages} {3865} (\bibinfo {year} {1996}{\natexlab{b}})}\BibitemShut {NoStop}%
\bibitem [{\citenamefont {Lu}\ and\ \citenamefont {Chen}(2012)}]{luMultiwfnMultifunctionalWavefunction2012}%
  \BibitemOpen
  \bibfield  {author} {\bibinfo {author} {\bibfnamefont {T.}~\bibnamefont {Lu}}\ and\ \bibinfo {author} {\bibfnamefont {F.}~\bibnamefont {Chen}},\ }\bibfield  {title} {\bibinfo {title} {Multiwfn: {{A}} multifunctional wavefunction analyzer},\ }\href {https://doi.org/10/cb7295} {\bibfield  {journal} {\bibinfo  {journal} {J. Comput. Chem.}\ }\textbf {\bibinfo {volume} {33}},\ \bibinfo {pages} {580} (\bibinfo {year} {2012})}\BibitemShut {NoStop}%
\bibitem [{\citenamefont {Liu}\ \emph {et~al.}(2020)\citenamefont {Liu}, \citenamefont {Lu},\ and\ \citenamefont {Chen}}]{liuSphybridizedAllcarboatomicRing2020}%
  \BibitemOpen
  \bibfield  {author} {\bibinfo {author} {\bibfnamefont {Z.}~\bibnamefont {Liu}}, \bibinfo {author} {\bibfnamefont {T.}~\bibnamefont {Lu}},\ and\ \bibinfo {author} {\bibfnamefont {Q.}~\bibnamefont {Chen}},\ }\bibfield  {title} {\bibinfo {title} {An sp-hybridized all-carboatomic ring, cyclo[18]carbon: {{Electronic}} structure, electronic spectrum, and optical nonlinearity},\ }\href {https://doi.org/10/gndmv6} {\bibfield  {journal} {\bibinfo  {journal} {Carbon}\ }\textbf {\bibinfo {volume} {165}},\ \bibinfo {pages} {461} (\bibinfo {year} {2020})}\BibitemShut {NoStop}%
\bibitem [{\citenamefont {Wolkin}\ \emph {et~al.}(1999)\citenamefont {Wolkin}, \citenamefont {Jorne}, \citenamefont {Fauchet}, \citenamefont {Allan},\ and\ \citenamefont {Delerue}}]{wolkinElectronicStatesLuminescence1999}%
  \BibitemOpen
  \bibfield  {author} {\bibinfo {author} {\bibfnamefont {M.~V.}\ \bibnamefont {Wolkin}}, \bibinfo {author} {\bibfnamefont {J.}~\bibnamefont {Jorne}}, \bibinfo {author} {\bibfnamefont {P.~M.}\ \bibnamefont {Fauchet}}, \bibinfo {author} {\bibfnamefont {G.}~\bibnamefont {Allan}},\ and\ \bibinfo {author} {\bibfnamefont {C.}~\bibnamefont {Delerue}},\ }\bibfield  {title} {\bibinfo {title} {Electronic {{States}} and {{Luminescence}} in {{Porous Silicon Quantum Dots}}: {{The Role}} of {{Oxygen}}},\ }\href {https://doi.org/10.1103/PhysRevLett.82.197} {\bibfield  {journal} {\bibinfo  {journal} {Phys. Rev. Lett.}\ }\textbf {\bibinfo {volume} {82}},\ \bibinfo {pages} {197} (\bibinfo {year} {1999})}\BibitemShut {NoStop}%
\bibitem [{\citenamefont {Canham}(1990)}]{canhamSiliconQuantumWire1990}%
  \BibitemOpen
  \bibfield  {author} {\bibinfo {author} {\bibfnamefont {L.~T.}\ \bibnamefont {Canham}},\ }\bibfield  {title} {\bibinfo {title} {Silicon quantum wire array fabrication by electrochemical and chemical dissolution of wafers},\ }\href {https://doi.org/10/bfr85g} {\bibfield  {journal} {\bibinfo  {journal} {Appl. Phys. Lett.}\ }\textbf {\bibinfo {volume} {57}},\ \bibinfo {pages} {1046} (\bibinfo {year} {1990})}\BibitemShut {NoStop}%
\bibitem [{\citenamefont {Lu}\ \emph {et~al.}(1995)\citenamefont {Lu}, \citenamefont {Lockwood},\ and\ \citenamefont {Baribeau}}]{luQuantumConfinementLight1995}%
  \BibitemOpen
  \bibfield  {author} {\bibinfo {author} {\bibfnamefont {Z.~H.}\ \bibnamefont {Lu}}, \bibinfo {author} {\bibfnamefont {D.~J.}\ \bibnamefont {Lockwood}},\ and\ \bibinfo {author} {\bibfnamefont {J.-M.}\ \bibnamefont {Baribeau}},\ }\bibfield  {title} {\bibinfo {title} {Quantum confinement and light emission in {{SiO2}}/{{Si}} superlattices},\ }\href {https://doi.org/10.1038/378258a0} {\bibfield  {journal} {\bibinfo  {journal} {Cah. Rev. The.}\ }\textbf {\bibinfo {volume} {378}},\ \bibinfo {pages} {258} (\bibinfo {year} {1995})}\BibitemShut {NoStop}%
\bibitem [{\citenamefont {Pavesi}\ \emph {et~al.}(2000)\citenamefont {Pavesi}, \citenamefont {Dal~Negro}, \citenamefont {Mazzoleni}, \citenamefont {Franz{\`o}},\ and\ \citenamefont {Priolo}}]{pavesiOpticalGainSilicon2000}%
  \BibitemOpen
  \bibfield  {author} {\bibinfo {author} {\bibfnamefont {L.}~\bibnamefont {Pavesi}}, \bibinfo {author} {\bibfnamefont {L.}~\bibnamefont {Dal~Negro}}, \bibinfo {author} {\bibfnamefont {C.}~\bibnamefont {Mazzoleni}}, \bibinfo {author} {\bibfnamefont {G.}~\bibnamefont {Franz{\`o}}},\ and\ \bibinfo {author} {\bibfnamefont {F.}~\bibnamefont {Priolo}},\ }\bibfield  {title} {\bibinfo {title} {Optical gain in silicon nanocrystals},\ }\href {https://doi.org/10/chbbr4} {\bibfield  {journal} {\bibinfo  {journal} {Cah. Rev. The.}\ }\textbf {\bibinfo {volume} {408}},\ \bibinfo {pages} {440} (\bibinfo {year} {2000})}\BibitemShut {NoStop}%
\bibitem [{\citenamefont {Almeida}\ \emph {et~al.}(2018)\citenamefont {Almeida}, \citenamefont {Goldoni}, \citenamefont {Akkerman}, \citenamefont {Dang}, \citenamefont {Khan}, \citenamefont {Marras}, \citenamefont {Moreels},\ and\ \citenamefont {Manna}}]{almeidaRoleAcidBase2018}%
  \BibitemOpen
  \bibfield  {author} {\bibinfo {author} {\bibfnamefont {G.}~\bibnamefont {Almeida}}, \bibinfo {author} {\bibfnamefont {L.}~\bibnamefont {Goldoni}}, \bibinfo {author} {\bibfnamefont {Q.}~\bibnamefont {Akkerman}}, \bibinfo {author} {\bibfnamefont {Z.}~\bibnamefont {Dang}}, \bibinfo {author} {\bibfnamefont {A.~H.}\ \bibnamefont {Khan}}, \bibinfo {author} {\bibfnamefont {S.}~\bibnamefont {Marras}}, \bibinfo {author} {\bibfnamefont {I.}~\bibnamefont {Moreels}},\ and\ \bibinfo {author} {\bibfnamefont {L.}~\bibnamefont {Manna}},\ }\bibfield  {title} {\bibinfo {title} {Role of {{Acid}}\textendash{{Base Equilibria}} in the {{Size}}, {{Shape}}, and {{Phase Control}} of {{Cesium Lead Bromide Nanocrystals}}},\ }\href {https://doi.org/10.1021/ACSnano.7b08357} {\bibfield  {journal} {\bibinfo  {journal} {ACS Nano}\ }\textbf {\bibinfo {volume} {12}},\ \bibinfo {pages} {1704} (\bibinfo {year} {2018})}\BibitemShut {NoStop}%
\bibitem [{\citenamefont {Bannwarth}\ and\ \citenamefont {Grimme}(2014)}]{bannwarthSimplifiedTimedependentDensity2014}%
  \BibitemOpen
  \bibfield  {author} {\bibinfo {author} {\bibfnamefont {C.}~\bibnamefont {Bannwarth}}\ and\ \bibinfo {author} {\bibfnamefont {S.}~\bibnamefont {Grimme}},\ }\bibfield  {title} {\bibinfo {title} {A simplified time-dependent density functional theory approach for electronic ultraviolet and circular dichroism spectra of very large molecules},\ }\href {https://doi.org/10/f6cfc6} {\bibfield  {journal} {\bibinfo  {journal} {Comput. Theor. Chem.}\ }\bibinfo {series} {Excited States: {{From}} Isolated Molecules to Complex Environments},\ \textbf {\bibinfo {volume} {1040--1041}},\ \bibinfo {pages} {45} (\bibinfo {year} {2014})}\BibitemShut {NoStop}%
\bibitem [{\citenamefont {Guidon}\ \emph {et~al.}(2010)\citenamefont {Guidon}, \citenamefont {Hutter},\ and\ \citenamefont {VandeVondele}}]{guidonAuxiliaryDensityMatrix2010}%
  \BibitemOpen
  \bibfield  {author} {\bibinfo {author} {\bibfnamefont {M.}~\bibnamefont {Guidon}}, \bibinfo {author} {\bibfnamefont {J.}~\bibnamefont {Hutter}},\ and\ \bibinfo {author} {\bibfnamefont {J.}~\bibnamefont {VandeVondele}},\ }\bibfield  {title} {\bibinfo {title} {Auxiliary {{Density Matrix Methods}} for {{Hartree}}-{{Fock Exchange Calculations}}},\ }\href {https://doi.org/10.1021/ct1002225} {\bibfield  {journal} {\bibinfo  {journal} {J. Chem. Theory Comput.}\ }\textbf {\bibinfo {volume} {6}},\ \bibinfo {pages} {2348} (\bibinfo {year} {2010})}\BibitemShut {NoStop}%
\bibitem [{\citenamefont {Locardi}\ \emph {et~al.}(2018)\citenamefont {Locardi}, \citenamefont {Cirignano}, \citenamefont {Baranov}, \citenamefont {Dang}, \citenamefont {Prato}, \citenamefont {Drago}, \citenamefont {Ferretti}, \citenamefont {Pinchetti}, \citenamefont {Fanciulli}, \citenamefont {Brovelli}, \citenamefont {De~Trizio},\ and\ \citenamefont {Manna}}]{locardiColloidalSynthesisDouble2018}%
  \BibitemOpen
  \bibfield  {author} {\bibinfo {author} {\bibfnamefont {F.}~\bibnamefont {Locardi}}, \bibinfo {author} {\bibfnamefont {M.}~\bibnamefont {Cirignano}}, \bibinfo {author} {\bibfnamefont {D.}~\bibnamefont {Baranov}}, \bibinfo {author} {\bibfnamefont {Z.}~\bibnamefont {Dang}}, \bibinfo {author} {\bibfnamefont {M.}~\bibnamefont {Prato}}, \bibinfo {author} {\bibfnamefont {F.}~\bibnamefont {Drago}}, \bibinfo {author} {\bibfnamefont {M.}~\bibnamefont {Ferretti}}, \bibinfo {author} {\bibfnamefont {V.}~\bibnamefont {Pinchetti}}, \bibinfo {author} {\bibfnamefont {M.}~\bibnamefont {Fanciulli}}, \bibinfo {author} {\bibfnamefont {S.}~\bibnamefont {Brovelli}}, \bibinfo {author} {\bibfnamefont {L.}~\bibnamefont {De~Trizio}},\ and\ \bibinfo {author} {\bibfnamefont {L.}~\bibnamefont {Manna}},\ }\bibfield  {title} {\bibinfo {title} {Colloidal {{Synthesis}} of {{Double Perovskite Cs2AgInCl6}} and {{Mn-Doped Cs2AgInCl6 Nanocrystals}}},\ }\href {https://doi.org/10.1021/jACS.8b07983} {\bibfield  {journal} {\bibinfo  {journal} {J. Am. Chem. Soc.}\ }\textbf {\bibinfo {volume} {140}},\ \bibinfo {pages} {12989} (\bibinfo {year} {2018})}\BibitemShut {NoStop}%
\bibitem [{\citenamefont {Manna}\ \emph {et~al.}(2019)\citenamefont {Manna}, \citenamefont {Das},\ and\ \citenamefont {Yella}}]{mannaTunableStableWhite2019}%
  \BibitemOpen
  \bibfield  {author} {\bibinfo {author} {\bibfnamefont {D.}~\bibnamefont {Manna}}, \bibinfo {author} {\bibfnamefont {T.~K.}\ \bibnamefont {Das}},\ and\ \bibinfo {author} {\bibfnamefont {A.}~\bibnamefont {Yella}},\ }\bibfield  {title} {\bibinfo {title} {Tunable and {{Stable White Light Emission}} in {{Bi3}}+-{{Alloyed Cs2AgInCl6 Double Perovskite Nanocrystals}}},\ }\href {https://doi.org/10.1021/ACS.chemmater.9b02973} {\bibfield  {journal} {\bibinfo  {journal} {Chem. Mater.}\ }\textbf {\bibinfo {volume} {31}},\ \bibinfo {pages} {10063} (\bibinfo {year} {2019})}\BibitemShut {NoStop}%
\bibitem [{\citenamefont {Shirahata}\ \emph {et~al.}(2020)\citenamefont {Shirahata}, \citenamefont {Nakamura}, \citenamefont {Inoue}, \citenamefont {Ghosh}, \citenamefont {Nemoto}, \citenamefont {Nemoto}, \citenamefont {Takeguchi}, \citenamefont {Masuda}, \citenamefont {Tanaka},\ and\ \citenamefont {Ozin}}]{shirahataEmergingAtomicEnergy2020a}%
  \BibitemOpen
  \bibfield  {author} {\bibinfo {author} {\bibfnamefont {N.}~\bibnamefont {Shirahata}}, \bibinfo {author} {\bibfnamefont {J.}~\bibnamefont {Nakamura}}, \bibinfo {author} {\bibfnamefont {J.-i.}\ \bibnamefont {Inoue}}, \bibinfo {author} {\bibfnamefont {B.}~\bibnamefont {Ghosh}}, \bibinfo {author} {\bibfnamefont {K.}~\bibnamefont {Nemoto}}, \bibinfo {author} {\bibfnamefont {Y.}~\bibnamefont {Nemoto}}, \bibinfo {author} {\bibfnamefont {M.}~\bibnamefont {Takeguchi}}, \bibinfo {author} {\bibfnamefont {Y.}~\bibnamefont {Masuda}}, \bibinfo {author} {\bibfnamefont {M.}~\bibnamefont {Tanaka}},\ and\ \bibinfo {author} {\bibfnamefont {G.~A.}\ \bibnamefont {Ozin}},\ }\bibfield  {title} {\bibinfo {title} {Emerging {{Atomic Energy Levels}} in {{Zero-Dimensional Silicon Quantum Dots}}},\ }\href {https://doi.org/10/ghdnjn} {\bibfield  {journal} {\bibinfo  {journal} {Nano Lett.}\ }\textbf {\bibinfo {volume} {20}},\ \bibinfo {pages} {1491} (\bibinfo {year} {2020})}\BibitemShut {NoStop}%
\bibitem [{\citenamefont {Tran}\ and\ \citenamefont {Blaha}(2009)}]{tranAccurateBandGaps2009}%
  \BibitemOpen
  \bibfield  {author} {\bibinfo {author} {\bibfnamefont {F.}~\bibnamefont {Tran}}\ and\ \bibinfo {author} {\bibfnamefont {P.}~\bibnamefont {Blaha}},\ }\bibfield  {title} {\bibinfo {title} {Accurate {{Band Gaps}} of {{Semiconductors}} and {{Insulators}} with a {{Semilocal Exchange-Correlation Potential}}},\ }\href {https://doi.org/10.1103/PhysRevLett.102.226401} {\bibfield  {journal} {\bibinfo  {journal} {Phys. Rev. Lett.}\ }\textbf {\bibinfo {volume} {102}},\ \bibinfo {pages} {226401} (\bibinfo {year} {2009})}\BibitemShut {NoStop}%
\end{thebibliography}%


%

\end{document}